# Benchmarking structure-based three-dimensional molecular generative models using GenBench3D: ligand conformation quality matters

*Benoit Baillif[1], Jason Cole[2], Patrick McCabe[2], Andreas Bender[1,*].*

[1] Yusuf Hamied Department of Chemistry, University of Cambridge, Lensfield Rd, Cambridge CB2 1EW, United Kingdom

[2] Cambridge Crystallographic Data Centre, 12 Union Road, Cambridge CB2 1EZ, United Kingdom

## Abstract

Three-dimensional (3D) deep molecular generative models offer the advantage of goal-directed generation based on 3D-dependent properties, such as binding affinity for structure-based design within binding pockets. Traditional benchmarks created to evaluate SMILES or molecular graphs generators, such as GuacaMol or MOSES, are limited to evaluate 3D generators as they do not assess the quality of the generated molecular conformation. In this work, we hence developed GenBench3D, which implements a new benchmark for models producing molecules within a binding pocket. Our main contribution is the Validity$_{3D}$ metric, evaluating the conformation quality using the likelihood of bond lengths and valence angles based on reference values observed in the Cambridge Structural Database. Absolute and relative (to native ligands) affinities were estimated using Vina, Glide and Gold PLP scores.

The LiGAN, 3D-SBDD, Pocket2Mol, TargetDiff, DiffSBDD and ResGen models were benchmarked. We show that only between 0% and 11% of generated molecules have valid conformations. Performing local relaxation of generated molecules in the pocket considerably improved the Validity$_{3D}$ for all models by a minimum increase of 40%. For LiGAN, 3D-SBDD, or TargetDiff, the set of valid relaxed molecules shows on average higher Vina score (i.e. worse) than the set of raw generated molecules, indicating that the binding affinity of raw generated molecules might be overestimated. Using the other scoring functions, that give higher importance to ligand strain, only yield improved scores when using valid relaxed molecules; Glide score shows poor (i.e., higher than 0) median scores on the set of raw molecules for most models, therefore representing a more suitable scoring function to use on raw molecules to detect ligand strain. Using valid relaxed molecules, TargetDiff and Pocket2Mol show better median Vina, Glide and Gold PLP scores than other models. We have publicly released GenBench3D on GitHub for broader use: https://github.com/bbaillif/genbench3d

## Introduction

Machine learning and deep learning models have gained popularity over the last decade for chemistry-related tasks such as prediction of molecular properties[1], or goal-directed molecular generation[2] using objective functions such as the Quantitative Estimate of Drug-likeness (QED), logP or docking scores for structure-based drug design[3]. These generative models offer an alternative to traditional virtual screening methods[4–6] to find new active molecules by exploring the chemical space, avoiding time and resource expenditure required to screen ultra-large chemical libraries. While popular generative models[3,7–9], such as REINVENT or GraphINVENT, produce SMILES or graph representations, requiring conformer generation to obtain an initial 3D structure for docking into protein pocket 3D structures[10], 3D generative models that produce molecules directly in a 3D conformation alleviates the need for separate conformation generation. Moreover, 3D generative models represent an opportunity to optimize molecules based on 3D objective functions, such as docking scores that depend on the molecule pose inside the binding pocket of interest[11,12].

There are two main types of 3D generative models: whole-molecule generation, relying on latent space decoding or diffusion models, and auto-regressive models, that add atoms or fragments one-by-one to an existing construct[11]. The first published model, namely the LiGAN model[13], uses a Variational Auto-Encoder (VAE) architecture to encode voxelised atomic densities of the protein pocket and ligand 3D structures in two separate latent spaces, and uses a decoder to output the atomic density of a generated molecule, whose bonds are inferred using OpenBabel[14]. Once the protein pocket of interest is encoded, generation of new molecules with LiGAN can be performed in two ways: either the ligand latent space can be sampled from a standard normal distribution with a zero mean and unit variance, referred to as prior sampling, or the ligand latent space can be sampled around the encoding of a native ligand (using a variability factor to control how similar the generated molecules are compared to the input ligand), referred to as posterior sampling. The 3D-SBDD[15] model was the first reported auto-regressive model for structure-based design. It builds a molecule by adding atoms one by one to an existing molecular construct (that can be randomly initialised in the pocket space) in 3D. It first uses a 3D graph neural network (GNN), referred to as the "Context Encoder", that encodes the protein and existing molecular construct environment for a given point in space. A second GNN, referred to as Spatial Classifier, gives a probability of finding a given atom type at the sampled position. The atom type and position are then sampled to obtain a new atom in the molecular construct until a stop signal is reached (i.e. the probability of adding a new atom is under a given threshold). OpenBabel is then used to infer bonds between these atoms. The Pocket2Mol[16] model builds on top of 3D-SBDD but uses vector-based neurons and geometric vector perceptrons for 3D information processing, and adds a bond prediction step. The TargetDiff[17] and DiffSBDD[18] models are both diffusion models, that start from an initial set of atoms whose number and initial types are sampled from a known distribution, then positions (and types in the case of TargetDiff) are refined to create potential new ligands in 3D inside the pocket. The ResGen[19] is a tuned version of Pocket2Mol that is faster by explicitly encoding fewer pocket atoms and parallelizing some parts of the generation.

In total, dozens of structure-based models have been reported[11], each with different methods to assess their performance, i.e., there is currently no unified benchmark that evaluates the key aspects of 3D generation. The MOSES[20] and GuacaMol[21] benchmarks are the most popular benchmarks for molecular generative models. They evaluate the validity of generated molecules as the fraction of molecules sanitised in RDKit[22], the uniqueness as the fraction of unique molecular graphs, the novelty as the fraction of molecules unseen during model training, the similarity to training set molecules based on molecular fingerprints, as well as computing properties distributions for molecular weight, logP, SAScore[23] and QED[24]. While these metrics are suitable for molecular graph generation (with or without 3D-specific scoring for optimisation), albeit limited in a real discovery setting[25,26], they perform no evaluation based on 3D atomic coordinates. A toy 3D generative model generating molecules with valid graphs unseen during training and with a similar property distribution to known dataset, but with all atomic coordinates set to [0,0,0] would show good results in GuacaMol or MOSES, but be useless in 3D structure-based design. These molecules would require conformation generation, commonly performed in virtual screening, but this would lose the main objective of 3D generative models to directly produce molecules along with their conformation for 3D-specific property optimisation.

There are two recent benchmarks for evaluation of 3D ligand structure in pockets: PoseBusters[27] (PB) and PoseCheck[28] (PC). The PB benchmark was designed to evaluate the output of deep learning methods for docking. It performs three checks: Chemical validity and consistency to ensure that the docked molecule is the same as the input molecule (e.g. no bond order or chirality was changed), intramolecular validity assessment using distance geometry included in RDKit[22], and intermolecular validity (i.e. ligand and protein) based on distance and volume. Based on the combination of these three checks, the authors concluded that these models generated up to 95% invalid poses, despite showing a Root-Mean-Square Deviation (RMSD) to the reference pose lower than 2 Å. However, the intramolecular and intermolecular validities have been computed with large margins compared to distance geometry bounds computed in RDKit: a molecule was counted as PB-valid if all distances were above 0.75 times their lower bound, and below 1.25 times their upper bound. While extreme cases might occur, a molecule with all bonds at either 0.75 lower bound or 1.25 upper bound would pass the check, while being energetically still highly unlikely. These cases are detected by the Energy ratio metric in PB, where 3% of the generated poses, albeit showing bond lengths within acceptable ranges, have energies with a ratio higher than 100 compared to the average energy of a set of 50 plausible conformers for the same molecule. The PC benchmark evaluated molecules produced by 3D generative models using the Vina re-docking RMSD, strain energy, interaction fingerprint similarity to training molecules, and steric clashes. PC detected that median re-docking RMSDs are higher than 2 Å for benchmarked models, suggesting that the docking algorithm did not find the same poses as the ones initially generated by the model. Moreover, while the re-docked poses showed on average half the number of steric clashes with the protein, they also showed lower interaction fingerprint similarity to known actives, indicating that protein-ligand geometry improved after redocking but led to different interactions, consistent with the higher re-docking RMSD.

While these benchmarks and original publications linked to models steer the evaluation of structure-based 3D generative models in the right direction, they are incomplete for assessing the objectives of 3D generation. These models should generate molecules with conformations having realistic geometries without the need for re-docking, the latter contradicting the very objective of structure-based 3D generation that extract insights from binding pocket positions to generate molecules in a good pose with favourable interactions. PB only incorporates both topological and geometric checks for deep docking models, starting from an input ligand molecular graph, while 3D generation as described in this work refers to de novo generation of new molecules directly in 3D space. PC only studies geometry with clashes and strain energy, and protein-ligand interactions, comparing raw and re-docked output. Both PB and PC do not include docking score evaluations, but they are commonly showed in the publications linked to the models. The main limitation is that docking scores are often evaluated on re-docked poses, hence not evaluating the actual output of the model. Regarding geometry, latest publications reported the Kullback-Leibler or Jensen-Shannon divergence between the distributions of bond lengths and valence angles for each atom or bond pair[16,17,19]. However, while able to compare values between different models, or models to a reference set, these values do not inform on the overall quality of the generated 3D structure (e.g., extreme bond lengths).

We implement in this benchmark the $Validity_{3D}$ metric previously proposed in our review[11]. While PB used RDKit distance geometry that rely on ideal values and a set of rules, we chose to rely on existing structural data to determine the quality of a molecular conformation. The Cambridge Structural Database (CSD)[29] contains more than a million curated crystal structures of organic and metal-organic molecules. The values for bond lengths, valence angles and torsion angles in the CSD for each pattern are greatly variable. Making use of this information, Mogul[30] uses the number of standard deviation away from the mean, namely the z-score, to evaluate the geometry of bond lengths and valence angles, with a z-score higher than 2 defining an unusual value, while any torsion angles where less than 5% of the distribution is within 10 degrees of the query value is unusual. However, as pointed out by McCabe and Cole[31], this might be misleading in the case of a distribution where the mean is away from the mode (or modes in the case of multimodal distributions). They argue that the likelihood of a geometrical value can be assessed relative to the mode of the distribution, and defined the q-value as the ratio between the likelihood of a query value over the maximum likelihood (i.e. likelihood at the mode). Taking the geometric mean of the values for all geometrical units of a conformation leads to a single value for the conformation, referred to as s-value, that estimates the global geometric validity of the conformation. As it is known that a ligand conformation can be strained upon binding[32], especially torsional strain to fit in the pocket, we decided to use a version of the s-value that only takes the bond lengths and valence angles. This s-value gives a normalized value (between 0 and 1) of the geometric validity, allowing the user to define their own threshold for validity (0.001 was chosen in this work). On top of the intramolecular validity, a generated molecule is 3D-valid if it does not have any intramolecular steric clash, and has no puckered (i.e., non-flat) aromatic ring.

In this work, we benchmark six structure-based 3D generative models using the designed Validity$_{3D}$ metric, which considers valid each conformation that has likely bond length and valence angles based on a reference 3D molecular structure source, and that has no intramolecular steric clash. We also estimated the affinity of raw generated conformations compared to the reference ligand for each pocket using three different scoring functions typically used for docking. Finally, we performed a local minimization (i.e. relaxation) for all generated poses and recomputed the benchmark metrics to observe if this step improved the generated conformation quality and its effect on estimated affinity. The suite of metrics used in this benchmark is implemented in the GenBench3D Python package, available on GitHub: https://github.com/bbaillif/genbench3d

## Methods

### Evaluation of molecular graph metrics

To stay consistent with the previous molecular generative model benchmarks, namely GuacaMol[21] and MOSES[20], their metrics were reimplemented in GenBench3D. All metrics used in GenBench3D are listed in Table 1. A molecule was considered to have a valid graph when the molecule was successfully parsed by RDKit[22] v.2023.6 with default sanitization, that standardizes some non-standard valence states, Kekulize aromatic rings, assigns radical electrons, sets aromaticity, conjugation, and hybridization, and finally checks valence correspondence. As described in the RDKit Cookbook[33], sanitization allows manipulation of molecules in RDKit by ensuring that the molecules are represented by octet-complete Lewis dot structures, and represents a good filter for invalid topological structures. For a given set of molecules generated by some method, the molecular graph validity $Validity_G$ is the fraction of molecules with valid graphs computed as indicated in Equation 1:

$$Validity_G = \frac{N_{valid\ G}}{N_{total}} \qquad (1)$$

where, $N_{valid\ G}$ is the number of generated molecules parsed with RDKit and $N_{total}$ is the total number of molecules generated by the evaluated method.

To check if a model did not generate the same molecules multiple times, the molecular graph uniqueness $Uniqueness_G(K)$ was measured as the fraction of unique molecular graphs among the first $K$ generated molecules with a valid graph, computed as indicated in Equation 2:

$$Uniqueness_G(K) = \frac{N_{unique\ G}}{K} \qquad (2)$$

where, $N_{unique\ G}$ is the number of unique canon SMILES (i.e. representing valid graphs). The MOSES benchmark recommends $K = 1,000$ and $K = 10,000$. In our case for structure-based generation, up to 10,000 molecules were generated with each model (100 for each target of a set of 100 targets), but in practice not all molecules were valid therefore $K$ was set to the number of total number of molecules with valid graphs generated by the model.

To check if the model generated new molecules that were not seen during training, the molecular graph novelty $Novelty_G$ was measured as the fraction of molecules with valid graphs absent from the training set, computed as indicated in Equation 3:

$$Novelty_G = \frac{N_{novel\ G}}{N_{valid\ G}} \qquad (3)$$

where, $N_{novel\ G}$ is the number of valid generated molecules with a graph absent from the training set.

To estimate the diversity of generated molecules, the molecular graph similarity $Sim(L_i, L_j)$ between each pair $L_i, L_j$ of generated molecules with valid graphs in the list of unique graphs $L$ was computed using the Tanimoto similarity of Morgan Fingerprints (implemented in RDKit[34]) with radius 3 and including chirality, which ranges between 0 (dissimilar) and 1 (identical). The average molecular graph diversity $AvDiv_G$ was computed as indicated in Equation 4:

$$AvDiv_G = \frac{\sum_{i=1}^{N_L} \sum_{j=i+1}^{N_L} 1 - Sim(L_i, L_j)}{\frac{(N_L - 1)N_L}{2}} \quad (4)$$

where, $L$ is the list of unique valid graphs, $N_L$ is the number of graphs in $L$, and $Sim$ is the Morgan Fingerprints Tanimoto similarity described in the previous paragraph. While the $Novelty_G$ assesses if generated molecules are not identical to those in the training set, the generated molecules can still be similar. To estimate the similarity between generated molecules and training molecules, the maximum Morgan Fingerprints Tanimoto similarity to molecules in the training set was computed for each generated molecule.

To assess the synthesizability of generated molecules, the SAScore[23], an estimation of the synthetic accessibility based on molecular complexity and fragment presence, was computed using the FreeSASA module in RDKit. To estimate solubility, the logarithm of the predicted octanol/water partition coefficient (logP) was computed using RDKit Crippen.MolLogP function. The distribution of ring sizes in generated molecules was also computed in RDKit using the GetRingInfo function.

Table 1: List of all metrics computed in GenBench3D. The number between round brackets after metric names refers to the corresponding Equation in the main text.

| Category | Metric | Definition |
|---|---|---|
| Molecular graph metrics | Validity$_G$ (1) | Fraction of molecules with valid graphs. A molecular graph is valid if it successfully sanitized by RDKit |
| | Uniqueness$_G$(K) (2) | Number of unique molecular graph (among the first K molecules generated) divided by K |
| | Novelty$_G$ (3) | Fraction of molecules with no identical graph in the training set |
| | AvDiv$_G$ (4) | Average diversity computed by average Tanimoto distance (1 - similarity) between Morgan Fingerprints radius 3 |
| | Ring proportion | Proportion of ring sizes observed in generated molecules |
| | MW | Molecular weight in g/mol |
| | SAScore | Synthetic Accessibility Score: estimation of the synthetic accessibility ranging between 0 (easy) and 10 (difficult) |
| | logP | Logarithm of the predicted octanol/water partition coefficient |
| | Maximum training similarity | Maximum Tanimoto Similarity of Morgan Fingerprints compared to molecules in the training set |
| Molecular conformation metrics | Validity$_{3D}$ (6) | Fraction of molecules with valid conformations. A conformation is valid if all bond lengths and valence angles are valid, and shows no intramolecular steric clash or puckered 5 or 6 membered aromatic rings |
| | Uniqueness$_{3D}$ (7) | Number of unique valid conformations, based on a Torsion Fingerprint Deviation (TFD) threshold, divided by the number of generated molecules. Only conformations of the molecules whose graph was generated multiple times are compared. |
| | Novelty$_{3D}$ (9) | Fraction of molecules with no similar conformation (based on TFD threshold) in the training set, among the molecules whose graph were seen in training. |
| | AvDiv$_{3D}$ (11) | Average diversity between conformers computed by average TFD, among molecules whose graph was generated multiple times. |
| | Strain energy | MMFF94s energy difference between raw and MMFF94s relaxed conformation after a maximum of 1000 relaxation step |
| Pocket-based metric | Steric clash | Computing potential clash between generated molecule and conditioning pocket |
| | Distance to centroid | Distance between the average atomic coordinates of the generated molecule and the native ligand |
| Binding affinity metrics | Vina score | Vina score computed in place |
| | Minimized Vina score | Vina score after performing Vina local minimization |
| | Glide score | Glide score computed in place |
| | Minimized Glide score | Glide score after performing Glide local minimization |
| | Gold PLP score | Gold PLP score computed in place |

## Evaluation of conformation quality metrics with the Validity3D

The quality of each generated molecular geometry was evaluated based on reference data observed in the Cambridge Structural Database (CSD)[29]. The CSD Drug subset[35], containing CSD refcodes for 8632 crystal structures from 785 unique drug molecules, constructed from InChI matching with the DrugBank 5.0 database[36], was obtained from the CSD_Drug_Subset.gcd file provided in the Supplementary Information of the original publication. The structures were retrieved using the CSD Python API[29] v.3.0.16, read into RDKit using a .mol file intermediate, and only the largest fragment was selected for each structure using the LargestFragmentChooser in RDKit.

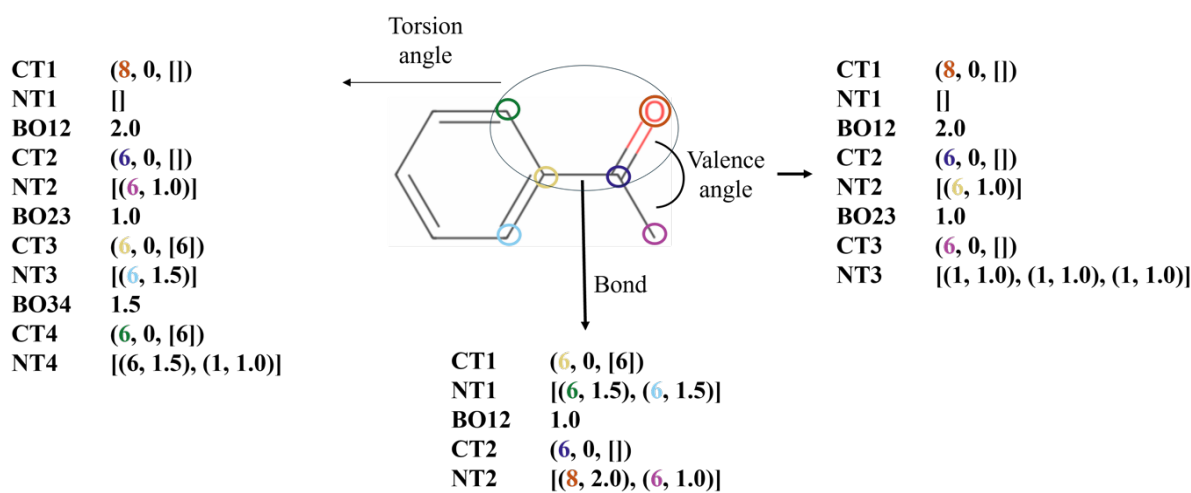

Figure 1: Examples of geometric pattern definitions for a bond, a valence angle, and a torsion angle in acetophenone. Each pattern is represented as a sequence of elements: central atom tuples CTX, with corresponding neighbourhood tuples NTX, separated by bond orders BOXY, where X and Y are central atom numbers. The atoms are pointed by coloured circles to identify corresponding values in the patterns. The hydrogens are not shown in the molecular graph but are indicated in the NTs.

Each bond length, valence angle, or torsion angle, referred to as a geometric pattern, was defined in this work by a pattern tuple that identifies the respective two, three or four atoms that form the geometric feature of interest, referred to as the 'central atom tuple' (CT) along with any neighbouring atom and bond type information. Examples of these patterns for the acetophenone are provided in Figure 1. A CT contains the information about the atomic number, formal charge, and a ring size tuple that stores the sizes of all the ring systems the central atom belongs to. Each central atom was associated with a neighbourhood tuple, referred to as NT, containing a list of neighbour atom tuples, each formed by the atomic number and the bond order to the corresponding central atom, for each neighbour atom except central atoms. The bond order, referred to as BO, was stored as a float, with the special case of 1.5 representing an aromatic bond (as identified in RDKit).

To allow for canonicalization of patterns, i.e., representing each geometric pattern with a single pattern tuple, an ordering system was setup, inspired by the Cahn–Ingold–Prelog priority rules. All tuples defined in the previous paragraph can be sorted with a sequence of keys, by descending order:

- Neighbour atom tuples in a NT are first sorted by atomic number, then bond order.
- CTs are first sorted by atomic number, then formal charge, then length of the ring size tuple, then pairwise comparison of the ring sizes.
- NTs are first sorted by number of neighbours, then pairwise comparison of their neighbour atom tuples.

Each geometric pattern was represented by a BO-separated sequence of combinations of CT with corresponding NT having sorted neighbour atom tuples. A series of comparisons was iterated: if a true condition was met, the geometric pattern sequence was reversed, and the series was stopped; the original sequence was kept if all conditions were false.

The series of comparisons for bond patterns was:
- First CT < second CT
- First NT < second NT

The series of comparisons for valence angle patterns was:
- First BO < second BO
- First CT < third CT
- First CT < third NT

The series of comparisons for torsion angle patterns was:
- Second CT < third CT
- First BO < third BO
- First CT < fourth CT
- Second NT < Third NT
- First NT < Fourth NT

For each bond, valence angle, and torsion angle pattern in the CSD Drug subset, the list of observed values was compiled in Python dictionaries where keys were the patterns defined as tuples.

To estimate the likelihood of geometrical values (bond length, valence angle, or torsion angle) with regards to known data, probability density functions (PDF) were approximated with kernel densities, that were fit for each pattern with at least 50 values, as described by McCabe et al[37]. The KernelDensity with Gaussian kernels as implemented in sklearn[38] v0.24.2 was used for bond length and valence angles, and an implementation of kernel density estimation with von Mises kernels for torsion angles was coded. The bandwidth (controlling the smoothing) for kernel densities was 0.01 for bonds, 1.0 for valence angles and 200 for torsion angles. Examples of fitted kernel density estimations for the pattern described in Figure 1 are shown in Figure S1.

Following McCabe and Cole[31], the q-value was used to estimate the normalized likelihood of a geometrical value by dividing the PDF value obtained with the kernel density by its maximum value (i.e. at the mode), as indicated in Equation 5:

$$q = \frac{PDF(x)}{\max(PDF)} \qquad (5)$$

The mode of each PDF was estimated by first uniformly sampling the PDF with 1,000 points ranging between 0.5 Å and 3.5 Å for bond lengths, between 0 and 180 degrees for valence angles and between -180 and 180 degrees for torsion angles, then using the minimize function from scipy.optimize with a Nelder-Mead optimization using the maximum sampled PDF as a starting point to find the maximum likelihood by minimizing the negative likelihood. Generalized versions of each pattern, where the NT were removed (the NTs of the two central atoms for bonds, and NTs of the first and last only or all central atoms for angle and torsions) were also computed, to account for the case when a pattern was encountered in a test setting but no kernel density was generated because it was seen less than 50 times in the reference dataset.

A bond or valence angle has in our work a valid geometry if its q-value was higher than a user-defined threshold, the default value being 0.001 in this work. The minimum and geometric mean of q-values per generated molecule were computed for the sets of all bonds, all valence angles, all torsions, the union of the sets of bonds and valence angles, and the union of the sets of bonds, valence angles and torsions. The geometric mean gives an idea of the overall 3D validity of the conformation based on CSD data, where a low q-value leads to a significant decrease of the geometric mean as opposed to computing the arithmetic mean, similar to the s-values in the work from McCabe and Cole[31]. On the other hand, the minimum q-value can directly detect if at least one geometry is invalid (i.e., q-value lower than threshold).

The presence of intramolecular steric clashes was also determined, defining a steric clash based on the van der Waals (vdW) radii, computed in RDKit with the GetRvdw function from the PeriodicTable object, for each pair of generated molecule atoms that were not involved in the same bond, valence angle or torsion angle (i.e. any steric clash between these uninvolved atoms would be detected by the invalid bond length and angle based on their q-value). The minimum distance between two heavy atoms was defined as the sum of their vdW radii multiplied by a user-defined factor, by default 0.75, to allow for a safety margin for the case of strained molecules, following PoseBusters[27]. The minimum distance between a heavy atom and a hydrogen atom was defined as the vdW radius of the heavy atom, while the minimum distance between two hydrogen atoms was the vdW radius of a hydrogen atom. A pair of atoms was considered as clashing if their distance was lower than the minimum distance.

Puckered aromatic rings were detected as follows[27]. For each 5-membered and 6-membered aromatic rings, eigenvectors of centred coordinates of its atoms were computed using singular value decomposition implemented in numpy[39], taking the last eigenvector as orthogonal to the plane, and projecting each atom coordinate to this eigenvector by dot product to obtain the perpendicular distance to the mean plane. Any ring with an atom with a distance to plane higher than a user-defined threshold (0.1 Å was used in accordance with PB) was considered puckered (i.e. invalid).

Based on all geometrical checks defined above, we introduce our main conformation quality metric in the GenBench3D benchmark, namely the Validity$_{3D}$, as introduced in our previous work[11]. A generated molecule was 3D-Valid, i.e., has a valid conformation, if all its bond lengths and valence angles were geometrically valid (based on the minimum q-value threshold), had no intramolecular steric clash (based on the multiplication factor applied to the vdW radii sum) and didn't show puckered aromatic rings (based on a maximum plane distance threshold). The Validity$_{3D}$ for a set of generated molecules was computed as indicated in Equation 6:

$$Validity_{3D} = \frac{N_{valid\ 3D}}{N_{valid\ G}} \tag{6}$$

where, $N_{valid\ 3D}$ is the number of 3D-valid molecules.

## Relaxation of Generated Poses

A constrained energy minimization (referred to as relaxation) was performed on each generated molecule in the pocket with the MMFF94s force field as implemented in RDKit. The pocket and generated molecule were merged in RDKit, the pocket atoms were fixed (i.e. they will not move during relaxation) and a distance constraint of 1 Å was set for each ligand heavy atom to avoid large ligand displacement, keeping a pose similar to the one generated by the model. A maximum of 1,000 minimization steps were performed.

## Similarity analysis of conformations using Torsion Fingerprint Deviation

The conformation difference between two 3D-valid conformers (i.e. conformations of the same molecular graph) was evaluated with the Torsion Fingerprint Deviation (TFD)[40] as implemented in RDKit, a metric ranging between 0 (no deviation) and 1 (maximal deviation) that computes the deviation between the values of corresponding torsions with a weight factor decreasing depending on the distance to the most central torsion. Two conformers were considered similar if they have a TFD lower than a user-defined threshold $d$; in this work, $d$ = 0.2 was used as recommended in the original publication to compare generated conformers to bioactive conformations. The TFD was chosen in preference to the Atomic Root-Mean-Square Deviation (ARMSD) because the latter is unbounded and is known to scale with the size of the molecule[41]. The following metrics were only computed on the set $\mathcal{M}$ of generated molecules for the set of graphs $\mathcal{G}_M$ that have been generated multiple times, i.e., containing at least two conformers for each molecular graph $g \in \mathcal{G}_M$.

Initialising an empty set $\mathcal{U}_d$ to store unique 3D-valid conformation, each generated 3D-valid molecule in $\mathcal{M}$ were iterated in the original order of generation, adding it to $\mathcal{U}_d$ if there was no similar conformer in $\mathcal{U}_d$, i.e., either it was the first time this molecular graph was generated, or there were already conformers in the set but the TFD between the generated molecule and existing conformers in $\mathcal{U}_d$ were all higher than $d$. To test if the model generates different conformers for the repeatedly generated molecular graph, the Uniqueness$_{3D}$ was computed as indicated in Equation 7:

$$Uniqueness_{3D}(d) = \frac{N_{\mathcal{U}_d}}{N_{\mathcal{M}}} \qquad (7)$$

where, $N_{\mathcal{U}_d}$ is the number of generated molecules in $\mathcal{U}_d$, and $N_{\mathcal{M}}$ is the number of generated molecules in $\mathcal{M}$. The Uniqueness$_{3D}$ can be rationalized by the size of $\mathcal{M}$ and $\mathcal{G}_M$, as it is expected for this metric to decrease if the size of $\mathcal{G}_M$ is close to 1 for high values of $\mathcal{M}$.

To assess if the model generates new conformers for graphs that were seen during training, the intersection set $\mathcal{G}_T$ between the training molecular graphs and generated molecular graphs was first identified. For each graph $g$ in $\mathcal{G}_T$, the corresponding generated molecules were added in the set $\mathcal{C}$, and the corresponding training molecules were added in the set $\mathcal{T}$. The Novelty$_{3D}$ was computed as indicated in Equation 9:

$$Novel_d(c, g) = \begin{cases} 1 \; if \; \min_{t \in \mathcal{T}_g}\bigl(TFD(c,t)\bigr) \geq d \\ 0 \; otherwise \end{cases} \qquad (8)$$

$$Novelty_{3D}(d) = \frac{\sum_{g \in \mathcal{G}_T} \sum_{c \in \mathcal{C}_g} Novel_d(c, g)}{N_{\mathcal{C}}} \qquad (9)$$

where, $Novel_d(c, g)$ is the novelty of a conformation $c$ with regards to all conformations $t$ in the set $\mathcal{T}_g$ of training conformations for the same molecular graph $g$ given the distance threshold $d$ as indicated in Equation 8, and $N_{\mathcal{C}}$ is the number of molecules in $\mathcal{C}$.

To evaluate the diversity of generated conformers for the same molecular graph, the inter-conformation deviation (ICD) was defined as the average TFD between its conformers, computed as indicated in Equation 10:

$$ICD(\mathcal{M}_g) = \frac{\sum_{i=1}^{N_{\mathcal{M}_g}} \sum_{j=i+1}^{N_{\mathcal{M}_g}} TFD\bigl(\mathcal{M}_{g_i}, \mathcal{M}_{g_j}\bigr)}{\frac{\bigl(N_{\mathcal{M}_g} - 1\bigr) N_{\mathcal{M}_g}}{2}} \qquad (10)$$

where, $\mathcal{M}_g$ is the list of generated valid 3D structures corresponding to a single molecular graph $g$ in $\mathcal{G}_M$ and $N_{\mathcal{M}_g}$ is the number of conformers in $\mathcal{M}_g$. To then evaluate the average diversity for the whole set of generated molecules, the AvDiv$_{3D}$ is the average ICD, computed as indicated in Equation 11:

$$AvDiv_{3D} = \frac{\sum_{g \in \mathcal{G}_M} ICD(\mathcal{M}_g)}{N_\mathcal{G}} \tag{11}$$

A high $AvDiv_{3D}$ (close to 1) indicates that the model generates different conformers for the same molecular graph, while a low $AvDiv_{3D}$ (close to 0) indicates that the model generates very similar conformations for the same molecular graph.

The strain energy was computed using the energy difference between the raw generated molecular conformation and the minimized conformation in vacuum (i.e. unconstrained by the pocket, thus being different to the relaxation described in the previous section) using the MMFF94s[42] force field as implemented in RDKit, with a maximum of 1,000 minimization steps.

### Evaluation of pocket-based metrics

Each protein was extracted from the CrossDocked[43] 1.1 dataset, processed with MDAnalysis[44] v2.6.1 to select only protein atoms (i.e., removing heterogens and waters), hydrogens were added with OpenBabel v3.1.1, and exported in a protonated protein pdb file. The pocket was selected with MDAnalysis, defined as the atoms of residues having at least one atom within 5 Å from any generated ligand atom.

The protein-ligand steric clashes were measured using the same steric clash criterion as the intramolecular steric clash computed in the $Validity_{3D}$, comparing each pocket atom to each generated molecule atom. While a model can generate 3D-valid molecules, a given generated molecule might not be adequate for structure-based drug design if it clashes with the protein.

The distance to centroid was measured with the Euclidean distance between the centroid (i.e. average of all atom coordinates) of the original ligand and the centroid of the generated molecule, computed with scipy v.1.11.2[45]. The distance to centroid should be close to 0 Å; a high distance to centroid, e.g., higher than 10 Å, indicates that the model generated the molecule outside of the pocket.

### Evaluation of binding affinity metrics

The binding affinities between the protein and the generated molecules were estimated using different scoring functions.

The Autodock Vina[46] score of each molecule was computed using the score function in the Vina Python package 1.2.4 (no docking was performed, to keep the input ligand fixed). The protonated protein pdb file produced for the pocket-based metrics was processed into a pdbqt file with the prepare_receptor binary from the ADFRsuite 1.0[47]. Each generated ligand was prepared with Meeko 0.5.0 [48] to obtain a pdbqt string. The pocket for scoring was defined with a centre being the centroid of the native ligand and a size border of 35 Å, a value large enough to account for very large ligands that might have been generated.

The Vina score minimization, usually performed in docking pose postprocessing, was also computed using the optimize function in the Vina Python package, optimizing the ligand position, orientation and value of the torsion angles to find the minimum score[49]. It is important to note that this score minimization is different from the relaxation performed with MMFF94s, as the latter minimize the force field energy, while Vina minimization optimizes the Vina score.

The Glide[50] score was computed by calling the command line interface from the Schrodinger software suite v2023-3. Each protonated protein pdb file was transformed in a mae structure using the structconvert binary from Maestro[51]. The mae file was prepared in a grid file required for Glide input using the Glide command line with the grid generation mode, setting the centre to the centroid of the native ligand and box size to the default with the inner box to 10x10x10 Å$^3$ and the outer box size adaptive to the size of the native ligand. The "inplace" score was used to score without docking the ligand. The minimized Glide score was also computed ('mininplace' option) to allow OPLS force field relaxation of the ligand with a maximum heavy atom displacement of 1 Å. Positive Glide scores were set to 0, as they represent extreme score values due to a very high intra-ligand strain term.

The Gold[52] PLP[53] score was computed using the CSD Python API v3.0.16. The protonated protein pdb file was given as input to the Docker object, and ligands were prepared using the LigandPreparation class. The binding site was defined by selecting protein atoms with a maximum distance of 10 Å from any ligand atom (BindingSiteFromLigand function).

Positive Glide and Vina scores, and negative Gold PLP were set to 0, as they represent outlying estimated affinity (i.e. the range of values of native ligands is negative for Glide and Vina, and positive for Gold PLP).

Relative scores to the native ligand for each target, i.e., subtracting the native ligand score to the absolute computed score, were also computed to assess if generated molecules have better estimated binding affinity compared to the known ligands. The fraction of molecules with better score, i.e., with negative Glide and Vina relative score, and positive Gold PLP relative score, was computed for each target for each model. The native ligand was also relaxed to compute the relative score for the set of valid relaxed generated molecules.

To test if there was a score difference before and after relaxation, and between pairs of models for the same set of generated molecules (raw, or valid and relaxed), the difference of median scores for each target for those comparisons was computed, and the Wilcoxon paired-sample test was used with the null hypothesis that the distribution of differences is symmetric about zero, i.e., that the median difference is zero. The Benjamini-Hochberg[54] multiple test correction for false positive rate was applied to adjust p-values from Wilcoxon tests compiling tests between models on one hand (15 tests), and compiling tests between generated molecule sets on the other hand. Using the approximation that the distribution of differences between the two set of samples is normal ("method=approx" is given as parameter to the wilcoxon function in scipy), the z-statistic of the test was computed. The effect size for each comparison was computed by dividing the z-statistic of the test by the square root of the total number of sample (i.e., two times the number of evaluated targets). An absolute effect size was considered small between 0.2 and 0.5, medium between 0.5 and 0.8, and large above 0.8.

### Benchmarking generative models on a CrossDocked test subset

Six 3D molecular generative models were benchmarked in this work: LiGAN[13], 3D-SBDD[15], Pocket2Mol[16], TargetDiff[17], DiffSBDD[18] and ResGen[19]. A quick description and GitHub links to the code of each model is provided in Table 2. The overall workflow of GenBench3D is pictured in Figure 2.

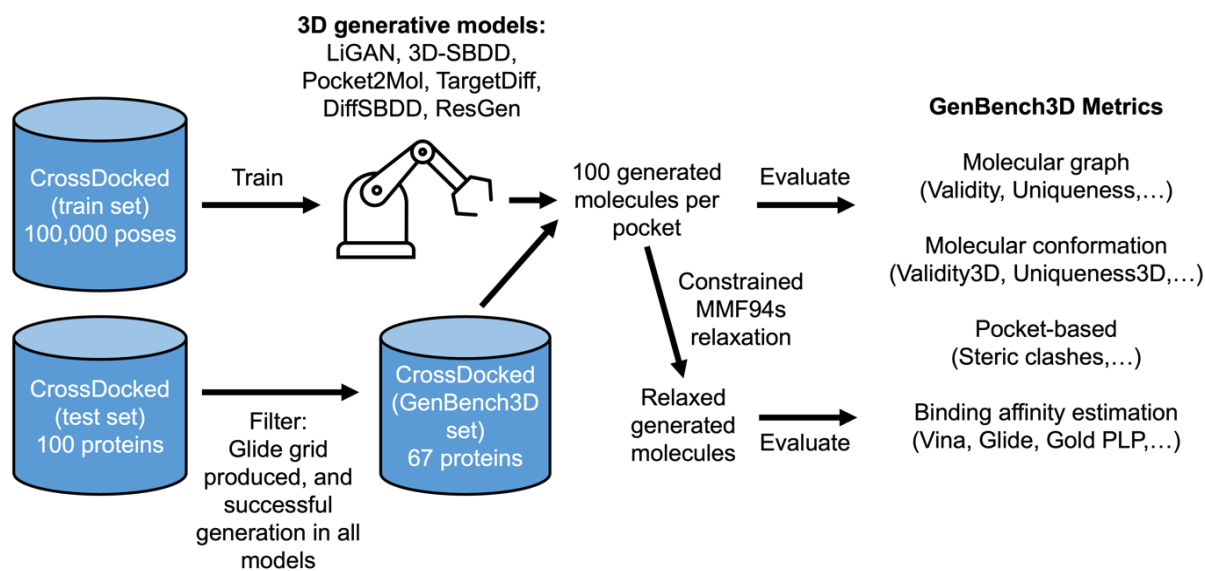

*Figure 2: GenBench3D workflow. CrossDocked was split between train and test set in previous works using a 30% sequence similarity threshold. Compared to these previous studies, we applied an additional filtering to the CrossDocked test set to include only protein pockets processed by the tools required for the molecule generation and metrics computation.*

For LiGAN, only the prior sampling, which is randomly sampling the ligand latent space from a standard normal distribution was performed; the posterior sampling, encoding a native ligand and sampling similar latent variables, was not included in this study as it would not be a fair comparison to other models that do not use any native ligand as input.
Code for the models and checkpoints of trained models with default parameters were downloaded from their respective GitHub pages. Snapshots of the code for the models run in this benchmark are provided in Supplementary Information for reproducibility. All models were trained by their respective authors with the CrossDocked[43] dataset version 1.1 using the same training set of 100,000 complexes, generating up to 100 molecules for each target in the test set for each model. The GenBench3D metrics described earlier were evaluated in this work using the same restricted set of 67 targets for which all models successfully generated molecules (e.g. ResGen failed for 20 targets) and Glide successfully generated a grid to compute score (from the original 100 targets in the CrossDocked test set).

The original 3D-SBDD, Pocket2Mol and ResGen generation script were tuned such that each generation path during the autoregressive sampling was a new path (compared to the original script that duplicates some favourable paths), to allow for more diverse molecules. Generated structures for TargetDiff were taken from the data supplied in the original publication, and molecules were generated with provided sampling scripts for all other models. Hydrogens were added to each generated molecule using RDKit for the evaluation of pocket-based and binding affinity metrics.

*Table 2: Generative models benchmarked in this study*

| Model name | Model type | Description | Code repository |
|---|---|---|---|
| LiGAN | Whole molecule encoding | Variational Auto-encoder encoding atom densities in a protein-ligand complex and decoding atom densities for ligands, conditioned by the protein pocket; requires additional atom and bond prediction step. | https://github.com/mattragoza/liGAN |
| 3D-SBDD | Autoregressive | The model uses a Context Encoder for protein + existing ligand atoms, and uses it in conjunction with a Spatial classifier to determine if a position is suitable to place an atom. Requires bond inference. | https://github.com/luost26/3D-Generative-SBDD |
| Pocket2Mol | Autoregressive | Improves on 3D-SBDD by adding a vector-based neuron + geometric vector perceptron architecture and predicting bonds during generation. | https://github.com/pengxingang/Pocket2Mol |
| TargetDiff | Whole molecule encoding | Diffusion model: an initial set of atoms is sampled and randomly placed in the pocket, and positions and atom types are refined through diffusion. | https://github.com/guanjq/targetdiff |
| DiffSBDD | Whole molecule encoding | Another diffusion model, but having a different formalism for atom type generation | https://github.com/arneschneuing/DiffSBDD |
| ResGen | Autoregressive | Improves on Pocket2Mol by only modelling alpha-carbon atoms and adding residue information (e.g. torsion angles) | https://github.com/HaotianZhangAI4Science/ResGen |

# Results and Discussion

## Evaluation of molecular graph metrics

In this work, we benchmarked six structure-based 3D molecular generative models, namely LiGAN, 3D-SBDD, Pocket2Mol, TargetDiff, DiffSBDD, and ResGen. We first evaluated molecular-graph-based performance metrics, similar to what is performed for SMILES or graph generators in GuacaMol or MOSES, the results of which are reported in Table S1. We describe the results in the "Evaluation of molecular graph metrics" section in the Supplementary Information. Overall, based on the molecular graph alone, generated molecules are valid, unique, novel, and diverse, have property values similar to those observed in the training molecules, and show low similarity to training molecules, also differing in the distribution of sizes of observed rings, with LiGAN, 3D-SBDD and DiffSBDD generating high fractions of rings with a size 3.

## Evaluation of conformation quality metrics on raw generated molecules

To assess the validity of generated 3D structures, we designed the Validity$_{3D}$ metric, representing the fraction of generated molecules with valid 3D conformation. We computed the Validity$_{3D}$ for all models, the CSD Drug Subset and the CrossDocked training set, and compiled the results in Table 3 and Figure 3. In ascending order, the Validity$_{3D}$ was 0.4% for DiffSBDD, 2% for LiGAN, 9% for Pocket2Mol, 10% for 3D-SBDD, and 11% for TargetDiff and ResGen. This result indicates that nearly 9 out of 10 molecules generated by every model has an unlikely geometry. The CrossDocked training set has a Validity3D of 48%, meaning that around half of training molecules show at least one geometric invalidity based on CSD data. This is an interesting result, as we would expect this dataset used to train 3D generative models to have valid 3D structures, i.e., with comparable geometry to those in the CSD. Other work also observed that the geometry of molecules in the CSD are different from those in the Protein Data Bank sourcing the CrossDocked dataset[55,56]. The CSD subset shows a Validity$_{3D}$ of 99%, suggesting that with our method we can expect a non-null fraction (e.g. around 1%) of conformations to be invalid due to at least one geometric outlier. Hence, the molecules generated by the models are mostly invalid based on their geometry, as opposed to the reference CSD data and to the CrossDocked training set.

.

*Table 3: Median conformation quality statistics of raw generated molecules*

| Molecule set | Validity$_{3D}$ | Minimum q-value | | | Geometric mean of q-values | | | Strain energy |
| --- | --- | --- | --- | --- | --- | --- | --- | --- |
| | | Bond | Angle | Bond and angle | Bond | Angle | Bond and angle | (kcal/mol) |
| LiGAN | 0.023 | 0.000 | 0.000 | 0.000 | 0.000 | 0.000 | 0.000 | 1508 |
| 3D_SBDD | 0.099 | 0.000 | 0.000 | 0.000 | 0.044 | 0.000 | 0.002 | 334 |
| Pocket2Mol | 0.094 | 0.000 | 0.000 | 0.000 | 0.010 | 0.059 | 0.020 | 133 |
| TargetDiff | 0.106 | 0.000 | 0.000 | 0.000 | 0.072 | 0.007 | 0.011 | 357 |
| DiffSBDD | 0.004 | 0.000 | 0.000 | 0.000 | 0.000 | 0.000 | 0.000 | 2006 |
| ResGen | 0.111 | 0.000 | 0.000 | 0.000 | 0.010 | 0.079 | 0.024 | 143 |
| CrossDocked (training) | 0.482 | 0.005 | 0.007 | 0.001 | 0.330 | 0.374 | 0.331 | 56 |
| CSD Drug | 0.989 | 0.241 | 0.230 | 0.143 | 0.725 | 0.728 | 0.720 | 120 |

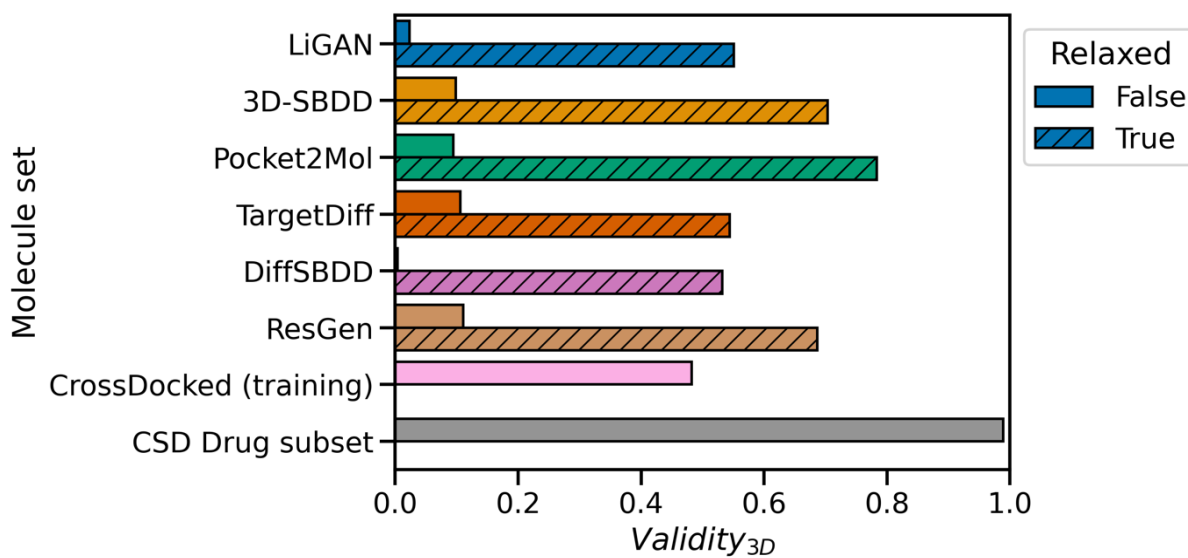

Figure 3: Validity$_{3D}$ of generated molecules before relaxation (open bars) and after relaxation (hatched bars)

While the Validity$_{3D}$ is a binary metric using a likelihood threshold, we next evaluated the extent of invalidity by computing the minima and geometric means of q-values (normalized likelihoods) for bond and valence angle patterns, reporting the medians in Table 3, and showing their distributions in Figure S2. We found that all models have a median minimum q-value of bond lengths and valence angles lower than the 0.001 threshold. This means that more than half the generated molecules have at least an invalid bond or angle, as reflected in the Validity$_{3D}$ metric that considers invalid any molecule that has a bond below the threshold. For comparison, the median minimum q-value for bonds and valence angles for the CrossDocked training set were 0.005 and 0.007 respectively, and for the CSD Drug Subset 0.241 and 0.230 respectively (high as expected since the kernel densities estimating the PDF were fit on this data distribution). It is worth noting that the median minimum q-value for the combination of bonds and valences angles, e.g., 0.143 for the CSD Drug Subset, is lower than their individual values because some molecules have more unlikely bonds while others have more unlikely angles. Additionally, we found that the geometric mean of bonds and valence angle q-values for generated molecules ranges from values lower than 0.001 for LiGAN to 0.024 for ResGen. Interestingly, the geometric mean of bond q-values was 0.072 for TargetDiff, showing a higher value than the geometric mean of angle q-values with 0.007, while for Pocket2Mol (or ResGen), the geometric mean of bond q-values was 0.010, lower than the geometric mean of angle q-values with 0.059. This result indicates that TargetDiff generate molecules with more likely bonds while Pocket2Mol generate molecules with more likely angles. These values are low compared to the CSD Drug subset with 0.720 or the CrossDocked training set with 0.331, indicating that a lot of q-values of geometric patterns in generated molecules are non-optimal (i.e. not close to 1). Overall, these results highlight a discrepancy between the distributions of bond lengths and valence angles observed in the CSD, and those observed in the generated molecules.

We further computed the median MMFF94s strain energy of the set of conformations generated by a model to estimate how strained the overall conformations are. As a reference, the median strain energy for the CrossDocked training set was 56 kcal/mol, and the median for the CSD Drug subset is 120 kcal/mol. These values are higher than the reported average upper bound of strain energy, e.g., up to 25 kcal/mol computed on 415 complexes in a study by Sitzmann et al.[57], but this upper bound was obtained on ligand conformations after relaxation. Moreover, MMFF94s was fit on data generated from high-quality quantum chemistry calculations (HF/6-31G*), therefore different from crystal structures that we can find in the CSD. We show that models generated molecules with median strain energy ranging between 133 kcal/mol for Pocket2Mol and 2006 kcal/mol for DiffSBDD, much higher than the median of the CrossDocked training set molecules.

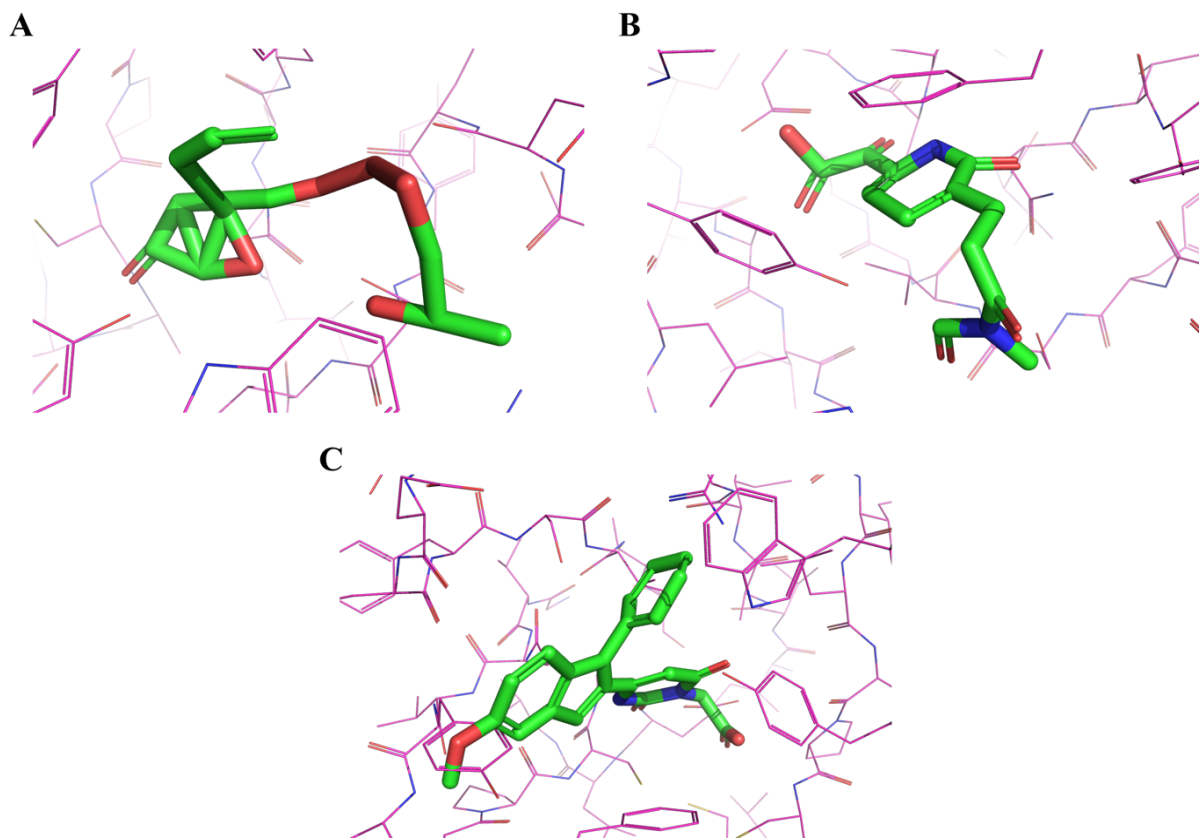

*Figure 4: Examples of molecules with invalid geometric structure generated in the cognate pocket for the protein with PDB identifier 2z3h. The generated molecules are shown in liquorice representation with green carbon atoms, while the protein pocket is shown in wire representation in purple carbon atoms. A: Molecule generated by LiGAN, showing extreme bond lengths and valence angles. B: Molecule generated by Pocket2Mol, showing a non-flat aromatic ring system. C: Molecule generated by ResGen, clashing with the protein pocket on the left-hand side.*

We next visually inspected some molecules with invalid structures: for instance, LiGAN generates molecules with extreme bond lengths, exceeding 2 Å for carbon-carbon bonds, invalid valence angles smaller than 45 degrees or larger than 140 degrees between non-triple bonds, as shown by the example in Figure 4A, while most models such as Pocket2Mol fail to generate flat aromatic ring systems, as shown by the example in Figure 4B.

Overall, all these results show that generated molecules have poor conformation quality when taking the 3D conformations as-is.

### Evaluation of conformation quality metrics on relaxed generated molecules

While the previous conformation quality analysis was performed on raw generated molecules, we next investigated whether these metrics could improve after minimizing the generated molecules within the binding pocket using the MMFF94s force field, allowing for a maximum deviation of 1 Å for heavy atoms and up to 1000 minimization steps. This minimization step will be referred to as relaxation throughout this manuscript. We examined the change of Validity$_{3D}$ after relaxation, reported in Table S3 and displayed in Figure 3. For all models, the relaxation drastically increased the Validity$_{3D}$: for instance, LiGAN increases from 0.02 to 0.55, Pocket2Mol increases from 0.09 to 0.78, and ResGen from 0.11 to 0.69. This relaxation pushes the Validity$_{3D}$ of all models on a par with the CrossDocked training set, indicating that it allows to reach conformations comparable to those obtained with docking studies that are commonly used in virtual screening.

Similarly, we analysed the minimum q-value of bond lengths and valence angle values, reporting the medians in Table S3 and the distributions in Figure S2. We show that the median of minimum q-values of bond lengths and valence angles are ranging between 0.002 and 0.016 after relaxation, while it was under 0.001 for all models before relaxation. We also examined the median geometric mean of q-values. On one hand, the median geometric means of q-values for bond length range between 0.37 for LiGAN to 0.56 for Pocket2Mol, much higher than the upper bound of 0.07 observed on non-relaxed molecules, indicating that the relaxation not only fixes invalid bonds, but also improves the likelihood of all bonds in general. This range is also slightly higher than the geometric mean of 0.33 for the CrossDocked training set, meaning that bond lengths in relaxed molecules are slightly more likely than those obtained via cross-docking. On the other hand, the median geometric mean of q-values for valence angles were lower than the median for bond lengths, ranging from 0.17 for LiGAN to 0.43 for Pocket2Mol. Overall, these results indicate that the force field relaxation fixes most invalid bonds and valence angles, to obtain geometries with similar likelihood to those in the training set.

We also analysed the median MMFF94s strain energy, which ranges from 10 kcal/mol for Pocket2Mol to 36 kcal/mol for DiffSBDD, compared to the range from 133 to 2006 kcal/mol observed without relaxation. These values are slightly lower than those obtained on the CrossDocked training set, but this is likely due to the difference between the force field optimization that works on multiple terms, i.e., not only bond and valence angle, but also torsion angle and electrostatic terms, while cross-docking was performed with smina that only minimizes the docking score by optimizing torsion angle values and ligand position in the pocket, hence using a ligand conformation with unrelaxed bonds and angles. Overall, these results show that relaxation drastically improved the conformation quality of generated molecules.

## Conformational difference analysis

After checking the validity of generated conformations, we next analysed conformation difference metrics on the set of 3D-valid relaxed generated molecules whose molecular graph were generated multiple times, namely the Uniqueness$_{3D}$, AvDiv$_{3D}$, and Novelty$_{3D}$ and reported the results in Figure 5 and Table S4. These metrics are based on the Torsion Fingerprint Deviation (TFD), that compares two different conformations for the same molecule based on the deviations of the values of their corresponding torsions. In this work, we considered conformations to be similar if they show a TFD lower than 0.2, as used in the original publication to compare generated conformers to bioactive conformations[40].

While the molecular graph uniqueness measures the fraction of unique graphs in generated molecules, the Uniqueness$_{3D}$ is the fraction of unique conformations among generated molecules whose graphs were generated multiple times (e.g. that have at least 2 conformers in the generated set). We can see that the Uniqueness$_{3D}$ ranges from 0.38 for Pocket2Mol (indicating that this model generated similar conformations for molecules having the same graph), to 0.80 for LiGAN. It is worth noting that CrossDocked has a low Uniqueness$_{3D}$ with a value of 0.10, which is expected since the same ligands were docked to multiple similar protein structures via cross-docking. Also, the CSD Drug Subset shows a low Uniqueness$_{3D}$ with a value of 0.20, indicating that in most cases when multiple crystal structures for the same ligand are available, they show similar conformations.

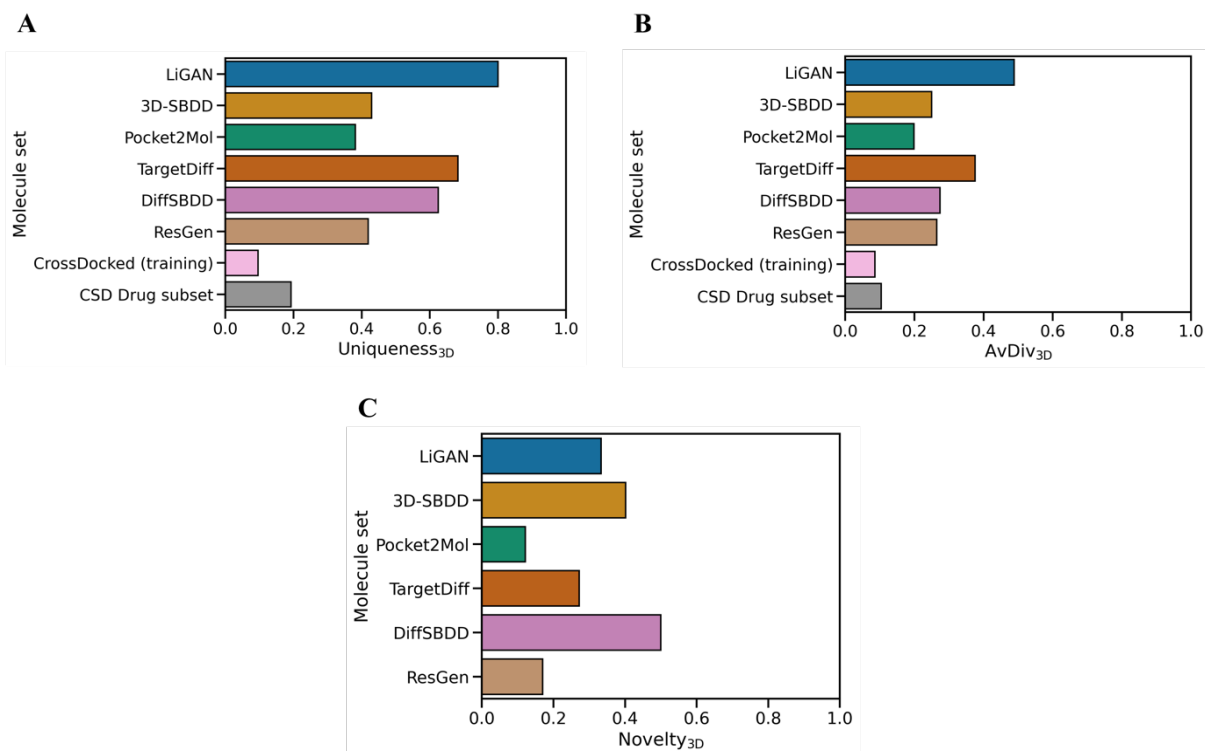

*Figure 5: Conformation difference metrics computed on the set of valid relaxed generated molecules whose graph have been generated multiple times. For CrossDocked (training) and the CSD Drug subset, no relaxation was performed for the Uniqueness$_{3D}$ (A) and AvDiv$_{3D}$ (B), and the Novelty$_{3D}$ (C) was not evaluated because they are reference sets.*

While the Uniqueness$_{3D}$ depends on a user-defined TFD threshold below which two conformations are classified as similar, the AvDiv$_{3D}$ uses the average TFD between conformations of the same molecular graph; we expect these two metrics to be correlated. We show that AvDiv$_{3D}$ ranges from 0.19 for Pocket2Mol to 0.49 for LiGAN, as reported in Figure 5B. A high value indicates that the model generates different conformations for the same molecule, which may be desirable in case a molecule has different binding modes. The AvDiv$_{3D}$ of the CrossDocked training set and CSD Drug Subset are around 0.10, indicating a lower diversity, which is expected for CrossDocked since the same ligand adopts similar conformations during cross-docking. Also, we can see that the Novelty$_{3D}$ range from 0.12 for Pocket2Mol to 0.50 for DiffSBDD, as reported in Figure 5C, suggesting that most relaxed generated conformations for the known molecular graphs were seen during training. Moreover, the number of molecules whose molecular graphs were generated multiple times is different between the models, and for the training set, limiting the direct comparison between these sets of molecules for the Uniqueness$_{3D}$ and AvDiv$_{3D}$.

This benchmark has been developed for structure-based generative models, but conformation difference might be important for other tasks such as distribution learning, with applications for conformation generation.

Pocket-based metrics analysis

We next analysed if generated molecules were clashing with the protein pocket, defining a clash by a ligand atom being too close to a protein atom, as these are unrealistic poses from a molecular perspective. Protein-ligand clashes is a failure mode of deep learning algorithms for molecular docking identified both by the PoseBusters[27] (PB) benchmark, and also identified for deep 3D molecular generators by the PoseCheck[28] (PC) benchmark. The fractions of generated molecules not clashing with the pocket before relaxation, shown in Figure 6, range from 0.97 for LiGAN to 0.28 for ResGen. The high steric clash fraction observed for ResGen can be explained by the pocket representation used by the model, that only explicitly encode backbone atoms and uses relative torsion angle information of residue atoms. ResGen does not detect clashes and can generate new atoms for the constructed ligand overlapping with pocket atoms, as shown in an example in Figure 4C, where the ligand clashes with the Tyrosine residue in the left-hand side of the figure. None of the models incorporate clash detection, hence the significant fraction of generated molecules clashing with the protein.

Next, we investigated the fraction of clashing molecules after ligand relaxation (i.e. minimizing the ligand within the pocket), shown in Figure 6, as we expect the clashes to reduce after this step. The fractions of non-clashing molecules were improved for all models after relaxation, increasing to a range of 0.98 for LiGAN to 0.81 for ResGen. It is important to note that the performed relaxation is constrained to a maximum deviation of each heavy atom to 1 Å, as a higher number might displace the molecule towards the outside of the pocket, losing the generated pose to achieve a relaxed state. We argue that if the generated molecule has a lot of clashes and has a high relaxation RMSD, the model failed at generating a suitable candidate that properly utilise pocket knowledge for generation.

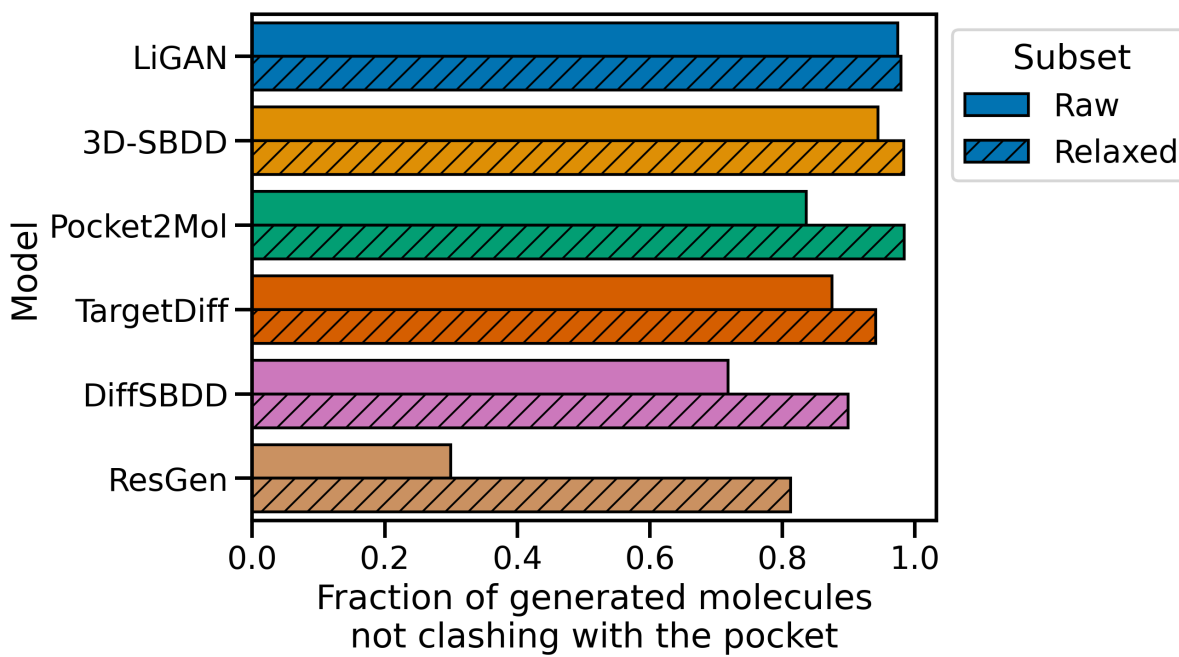

*Figure 6: Fractions of molecules generated by different models not clashing with the conditioning pocket for the sets of raw (open bars) and valid relaxed (hatched bars) generated molecules.*

We next analysed the fraction of molecules that were generated outside of the binding pocket, by computing the distance between the generated molecule centroid and native ligand centroid, and labelling a distance higher than 10 Å as "out of the pocket". The result of this analysis is shown in Table S5. We show that except for DiffSBDD that showed an "out of the pocket" fraction of 0.03, all models had fractions lower than 0.002. An example is shown in Figure S3, where the molecule generated by DiffSBDD is located at the surface of the pocket, while we expect the generation to occur close to the native ligand inside the pocket. The relaxation did not significantly change the fractions of generated molecules inside the pocket, as expected since a maximum heavy atom displacement of 1 Å was setup. Overall, these results show that almost the totality of generated molecules are inside the binding pocket for all models except DiffSBDD that show a higher fraction than the others.

### Absolute pose binding affinity estimation on raw generated molecules

The objective of structure-based deep 3D molecular generation is to generate new active candidates in the binding pocket of interest. In order to estimate the binding affinity of generated molecules in each pocket in the test set, we next computed five different binding affinity scores commonly used in docking algorithms, namely the Vina "inplace" score, Glide

"inplace" score and Gold PLP score in place, and a local minimization of Vina and Glide ("mininplace") scores that slightly displace the ligand to optimize the score. These scores are computed on two sets of molecules: the complete set of all non-relaxed generated molecules (i.e., whether they are 3D-valid or not), and the filtered set of 3D-valid among relaxed molecules, as this last set represents a more accurate situation similar to poses that docking algorithms can produce.

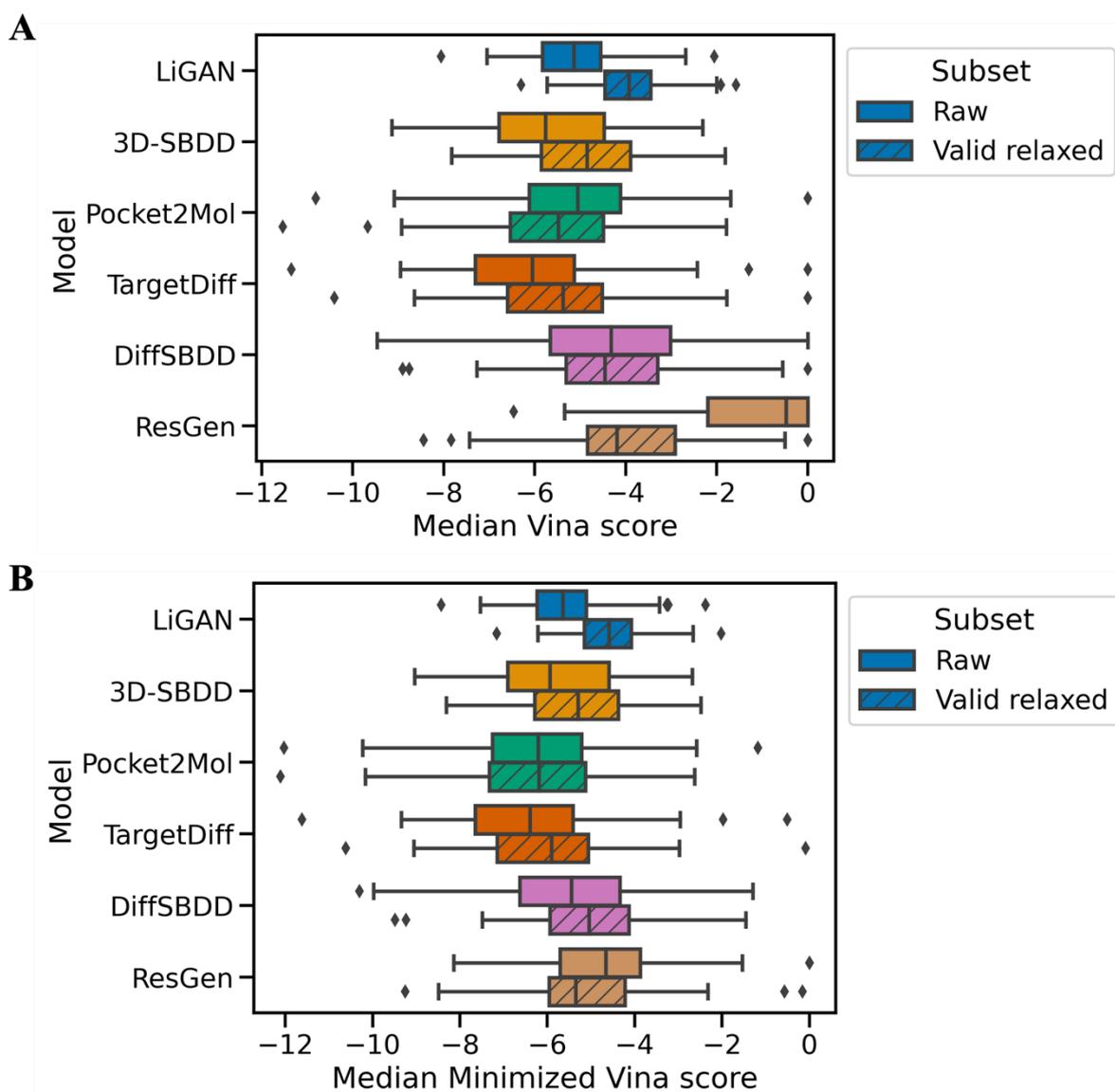

Figure 7: Distributions of median Vina (A) and minimized Vina (B) scores for the sets of raw (open bars) and valid relaxed (hatched bars) molecules. A lower Vina score indicates a better estimated binding affinity.

We next analysed the difference in median Vina scores distributions, representing the distribution of median Vina scores for each model in Figure 7, and analysed the paired difference with the Wilcoxon test with the null hypothesis that the median difference is zero, reporting p-values and effect sizes in Table S6. Using the medians of Vina score to assess the raw poses of generated molecules for each pocket, the median of pocket medians ranged between -6.05 for TargetDiff to -0.47 for ResGen, as shown in Figure 7A. We found that 3D-SBDD with a median of -5.76 and TargetDiff have significantly better (i.e., lower Vina scores) than the other models with p-values lower than 1e-03 and absolute effect sizes ranging from 0.39 (small) to 0.61 (medium) as indicated in Table S6. ResGen generated molecules with significantly worse Vina score than other models (and close to 0), which is a surprising result considering that it is the most recent model among those benchmarked in this work, with the original publication reporting that ResGen outperformed Pocket2Mol based on top-10 mean Vina score[19], with a mean of -9.08 versus -8.73. The main difference between their study and ours is that they used the Vina score after re-docking. The latter contradicts the objective of direct 3D generation in the pocket and optimize the position and geometry of the ligand to best fit the pocket, hence leading to lower (i.e. better) Vina scores, while the docked pose might be different from the generated one, as shown by PC[28], another benchmark for structure-based 3D molecular generative models. The Vina scores for ResGen raw generated molecules are strongly positively correlated to the number of steric clashes (Spearman coefficient = 0.85), as displayed in Figure S4, suggesting that the high Vina scores for this model might be caused by the high number of clashes. For the other models, all molecules having more than 5 clashes obtained a positive Vina score.

After local optimization of the raw generated molecules to minimize the Vina score, the median improved (i.e., decreased) for all models: for instance, ResGen decreases from -0.47 to -4.66, and Pocket2Mol from -5.05 to -6.21, as shown by the difference between Figure 7A and Figure 7B. With these minimized Vina scores, Pocket2Mol and TargetDiff (with a median of -6.39) both outperform the other models with p-values lower than 2e-03 and absolute effect sizes ranging from 0.29 (small) to 0.60 (medium), as shown in Table S6. Using the minimized Vina score, Pocket2Mol with a median of -6.21 is better than 3D-SBDD with a median of -5.94 using the minimized Vina score (p-value < 1.0e-03, absolute effect size = 0.29), while using the Vina score inplace, 3D-SBDD with a median of -5.72 is better than Pocket2Mol with a median of -5.06. These statistically significant Vina score difference might seem low (i.e. with small effect size), but a two-fold difference in dissociation constant (evaluated in binding assays) represents an estimated 0.4 Vina score difference (1.3 difference for a ten-fold dissociation constant increase), as explained in the section "Relationship between equilibrium constant and estimated binding free energy" in Supplementary Information. These results indicate that performing a slight change in the pose to optimize the score leads to a different ranking of the best performing models: TargetDiff and 3D-SBDD show the best performances before optimization while TargetDiff and Pocket2Mol show the best performances after optimization.

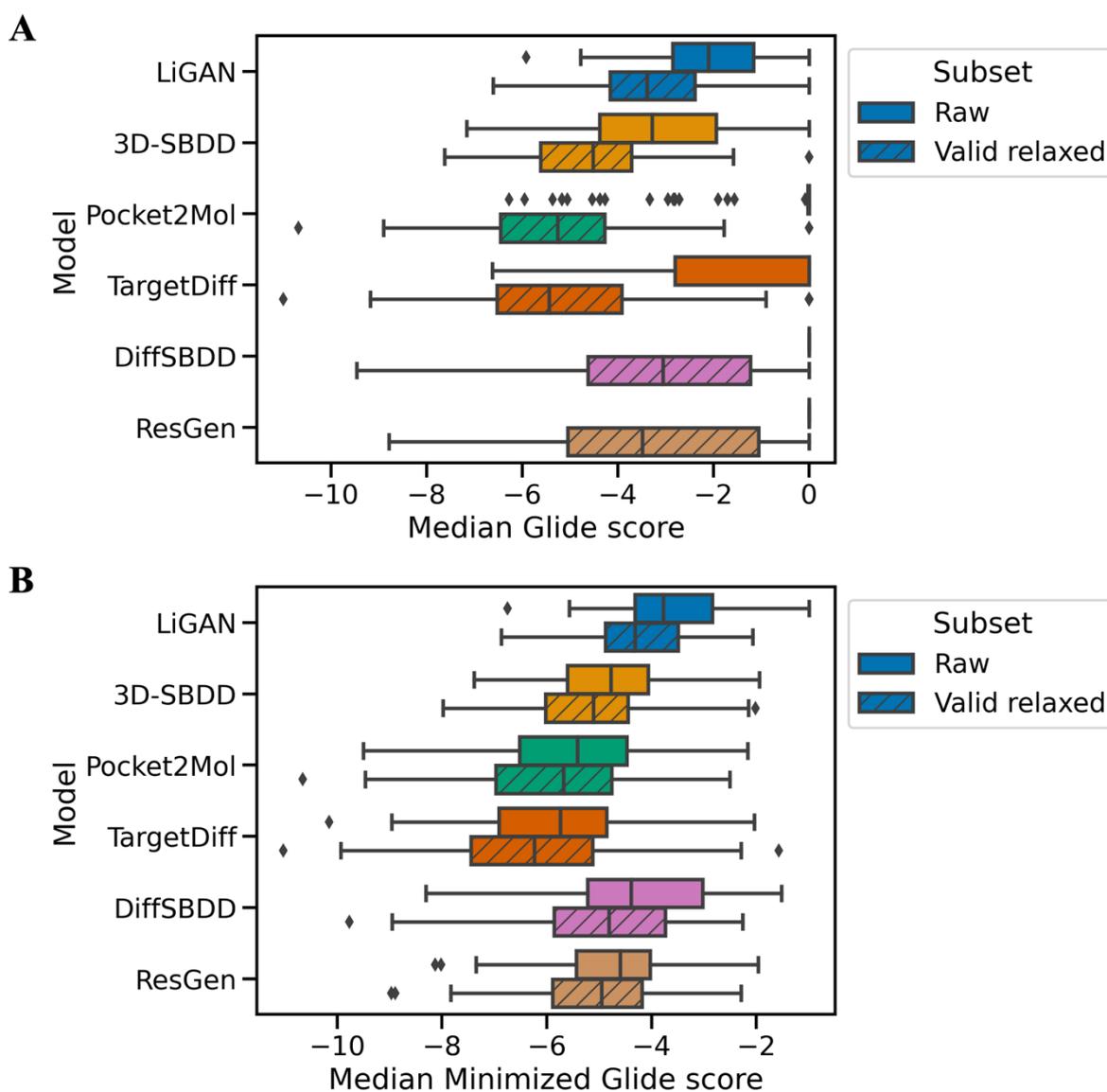

*Figure 8: Distributions of median Glide (A) and minimized Glide (B) scores for the sets of raw (open bars) and valid relaxed (hatched bars) molecules. A lower Glide score indicates a better estimated binding affinity.*

Using the median Glide scores, whose distributions are pictured in Figure 8A, 3D-SBDD shows significantly lower median scores, with a median of -3.28, compared to a range of -2.10 for LiGAN to 0 for all other models, with p-values lower than 1e-03 and absolute effect size ranging from 0.42 to 0.56 as shown in Table S6 (Glide score were bounded to 0 to avoid extreme values caused by the intramolecular strain energy term in the scoring function). Minimizing the Glide score (the "mininplace" option optimizes ligand position with maximum heavy atom displacement of 1 Å) improves the median score (like the Vina minimization), for instance for 3D-SBDD from -3.21 to -4.78 or for TargetDiff from 0.0 to -5.74, as shown in Figure 8B; Pocket2Mol, with a median of -5.41, and TargetDiff both outperform all other models, with p-values lower than 1e-03 and medium absolute effect sizes ranging from 0.52 to 0.61 (Table S6).

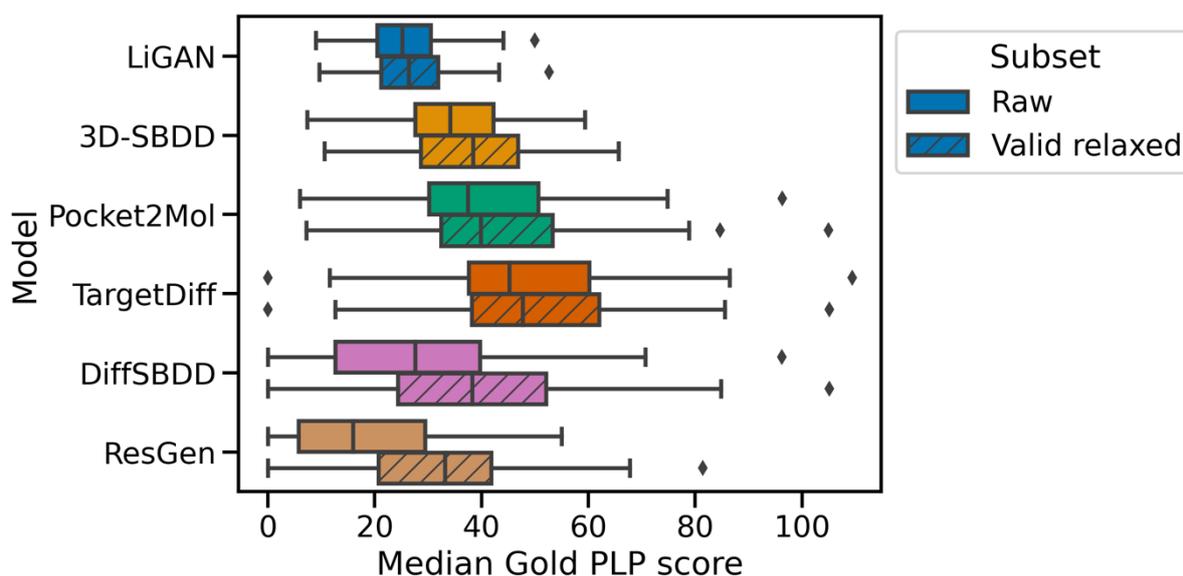

*Figure 9: Distributions of median Gold PLP scores for the sets of raw (open bars) and valid relaxed (hatched bars) molecules. A higher Gold PLP score indicates a better estimated binding affinity.*

Using the Gold PLP score, TargetDiff outperformed all other models with a median of 45.22 compared to a range of 15.90 for ResGen to 37.47 for Pocket2Mol, with p-values lower than 1e-03 and medium absolute effect sizes ranging from 0.52 to 0.60, as shown in the distributions in Figure 9 and p-values with effect sizes in Table S6.

Based on these results on raw poses, we show that the different scoring functions do not yield the same ranking for the models based on the median scores: while TargetDiff is among the best using Vina, minimized Glide and Gold PLP scores, 3D-SBDD is better when evaluating molecules with the Glide score inplace. This is due to the different scoring terms in each function, that weight differently the intramolecular ligand strain and steric clashes. Glide gives a positive score to invalid molecules because the intramolecular ligand strain term is more important, compared to the one of Vina which gives better score to models that show the highest rate of invalid molecules.

## Absolute pose binding affinity estimation on 3D-valid relaxed generated molecules

Since we showed in the previous section that scoring functions do not consistently give poor scores to geometrically invalid molecules, and that relaxation improved the validity of generated molecules, we next re-evaluated the scores on the set of 3D-valid relaxed molecules to have an analysis closer to what can be done in practice (i.e. with docking that already have likely molecular conformation as input), showing their distributions in Figure 7, Figure 8 and Figure 9. We also reported the p-values and effect size of the Wilcoxon test performed on the difference between these the set of raw (i.e. un-relaxed) molecules and the set of valid relaxed molecule for each model in Table S7.

Besides the combinations of DiffSBDD with Vina inplace scores, and LiGAN and TargetDiff with the Gold PLP score, there is a significant difference between the median of score medians of the set of raw generated molecules and the set of valid relaxed molecules, with p-values lower than 1e-03 and absolute effect sizes ranging between 0.40 and 0.61, as reported in Table S7. Besides from the Vina inplace scoring showing significantly increased scores indicating worse estimated binding affinity, e.g., with LiGAN going from a median of -5.14 to -3.93, all other scoring function and model combinations improved binding affinity estimation after relaxation. For instance, using Glide inplace, all models show decreased (i.e. better) scores after relaxation, with for instance Pocket2Mol, TargetDiff, DiffSBDD and ResGen that all had a median Glide score of 0 before relaxation, improved to -5.26, -5.43, -3.06 and -3.49 respectively, and ResGen improving from a median Gold PLP score of 15.90 before relaxation to 33.11 after relaxation. Comparing the scores of valid relaxed molecules between pairs of models, we found that Pocket2Mol and TargetDiff show significantly better estimated binding affinity than all models with p-values < 1e-03 and absolute effect sizes ranging between 0.24 (small) and 0.60 (medium) as indicated in Table S6.

Overall, the set of valid relaxed molecules show better scores than the set of raw molecules, which is expected since geometrical invalidities were identified earlier as cause of poor scores (i.e. intramolecular ligand strain for Glide, clashes for Vina). However, it is interesting to note that there are few models for which the Vina score does not improve on this refined set of molecules. This is concerning, because it means that using the Vina score on invalid molecules might lead to misleading conclusions on binding affinity estimation: 3D-SBDD is among the models producing the most active molecules using Vina on raw molecules, while it falls below Pocket2Mol after relaxation.

### Pose binding affinity estimation relative to native ligands

We next aimed to put predicted binding affinities into context, and hence compared each score to their reference cognate ligand (referred to as native) score for the same pocket, and computed the fraction of better scores (i.e. lower for Vina and Glide, higher for Gold PLP). This metric is similar to what has been done in the original publications linked to the benchmarked models with the High affinity measure[16,17], with the exception that the latter was computed on re-docked pose.

We found that models produce median fractions of molecules with better scores than the reference ligand ranging from 0% for ResGen to 20% for TargetDiff when using Vina inplace, from 0% for ResGen and DiffSBDD to 21% for 3D-SBDD when using Glide inplace, and from 0% for ResGen to 27% for TargetDiff when using Gold PLP score, as shown in Figure S5, Figure S6 and Figure S7. It is worth noting that the distribution of fractions for each model is very variable: for instance, TargetDiff that shows significantly higher median fractions than LiGAN, 3D-SBDD, DiffSBDD and ResGen using Vina and Gold PLP score (as indicated in Table S8), have fractions of generated molecules with better scores than native ranging from 0% to 100% depending on the target, indicating that the 3D generative model does not consistently generate better molecules for all targets, therefore not guaranteeing the usefulness of structure-based generative models.

Overall, these results indicate that even though most generated molecules have conformations with low likelihood, they still obtained better scores than the native ligand poses from CrossDocked, i.e., that were produced from cross-docking with smina[58], emphasizing the failure of scoring functions to give poor scores for these unrealistic conformations. In structure-based virtual screening, conformation generation sets realistic bond lengths, valence angles and torsion angles, and therefore the detection of unlikely geometry values does not have to be implemented in scoring functions, but this needs to be checked in the case of deep molecular generation in its current state.

We next analysed the change in the median fraction of generated molecules with better score than native on the set of valid relaxed molecules, compared to the set of raw molecules, indicating the p-values and absolute effect sizes in Table S9. For LiGAN, the median fraction decreased from 8% for raw molecules to 2% on the set of valid relaxed molecules using Vina and decreased from 5% to 0% when using Glide. For DiffSBDD and ResGen, the median fractions are higher after relaxation using all three scoring functions. For Pocket2Mol, the median fraction increased using Vina, but is similar with the other two scoring functions. The median fractions are similar for all other models before and after relaxation. When comparing the median fractions between models after relaxation, TargetDiff and Pocket2Mol show significantly higher median fractions than other models when using Vina and Glide, and only TargetDiff shows a significantly higher median for Gold PLP, as indicated in Table S8. Overall, the conclusions of the analysis on the absolute scores and fractions of generated molecules with better score than native are similar, suggesting that TargetDiff and Pocket2Mol generate molecules with better scores.

## Limitations of the benchmark

While the current analysis gives useful insights on the choice of current models and provides guidance on target metrics to sanity-check the structural quality of molecules generated with new models, there are still some limitations. The analysis was only made on 100 molecules per target, while in an actual drug discovery project it is desirable to generate many more candidate molecules to compete with traditional virtual screening. Virtual screening was not used as a baseline in this comparison because it works on a different time scale (i.e. in the order of seconds to dock a single molecule from a library containing thousands of molecules) and is parallelizable (scaling with CPU cores availability) while deep generative models exploit GPUs to generate multiple good molecules in the order of seconds but are only parallelizable to some extent (the larger the combined GPUs memory, the more nodes for diffusion models, and the more paths explored for autoregressive models). Virtual screening is expected to find better molecules in the long run, and has the advantage of using conformation generation that ensures that bond lengths and valence angles are in usual ranges, while benchmarked models required ligand relaxation to obtain 3D-valid poses. Moreover, the dataset used to train the models show different geometries compared to the CSD, as shown by the majority of ligands not being 3D-valid, hence it is not expected that the model extracts CSD-like ligand geometry without using transfer learning on additional CSD data[59].

## Conclusion

In this work, we defined the GenBench3D benchmark for 3D molecular generative models, that produce molecules along with their pose directly in the pocket, with the objective to generate molecules with good interaction profiles resulting in high binding affinity. To this date, these models have been incompletely evaluated, ignoring conformation validity, and using docking scores after re-docking, which contradicts the mere objective of generation within the pocket. Our main contribution is the implementation of the Validity$_{3D}$ metric described in a previous publication[11], that uses known distribution of geometry patterns in the CSD, and checks ring flatness and steric clashes to assess the realism of generated molecules. We benchmarked six structure-based molecular generative models and showed that they generated mostly molecules with invalid geometries. Relaxing the ligand in the pocket led to improvement of geometric likelihood, reaching Validity$_{3D}$ close to the training set molecules. Valid and relaxed molecules show significantly different binding affinity estimation with scoring functions used by docking algorithms, indicating that that these scoring functions should be used with caution, and justifying the need of using different functions to assess potential binding affinity between generated molecules and the protein pocket. GlideScore seems to be the best scoring function to use on unminimized molecules, giving poor values to invalid molecules. Based on all scoring functions, TargetDiff obtain better median scores compared to more recent models such as DiffSBDD or ResGen, and obtain close results to Pocket2Mol. However, it is worth mentioning that Pocket2Mol generated smaller molecules, and considering that scores scale with the molecule size, this model shows better ligand efficiency; it also produced molecules with better estimated synthetic accessibility and drug-likeness. We therefore recommend using both TargetDiff and Pocket2Mol to generate new molecules for their improved capacity to generate molecule with better estimated binding affinity.

# Abbreviations

3D: Three-dimensional

ARMSD: Atomic Root-Mean-Square Deviation

BO: Bond Order

CSD: Cambridge Structural Database

CT: Central atom Tuple

GNN: Graph Neural Network

NT: Neighbourhood Tuple

PB: PoseBusters

PC: PoseCheck

PDF: Probability Density Function

QED: Quantitative Estimate of Drug-likeness

RMSD: Root-Mean-Square Deviation

TFD: Torsion Fingerprint Deviation

SAScore: Synthetic Accessibility Score

VAE: Variational Auto-Encoder

vdW: van der Waals

# Acknowledgment

The authors would like to thank the University of Cambridge and the Cambridge Crystallographic Data Centre for funding this work.

# Author Contributions

B.B.: Conceptualization, Methodology, Software, Writing – Original Draft. J.C., P.M., and A.B.: Supervision, Writing – Reviewing and Editing.

# Funding

This work has been funded by the University of Cambridge and the Cambridge Crystallographic Data Centre.

## Data availability

The GenBench3D code is publicly available on GitHub:

https://github.com/bbaillif/genbench3d

The generated molecules for each model, and their relaxed counterparts are available on figshare : https://figshare.com/articles/dataset/Data_for_Benchmarking_structure-based_3D_generative_models_with_GenBench3D/26139496

The snapshots of code for models evaluated in this work are available on figshare:

https://figshare.com/articles/software/Snapshots_for_models_evaluated_in_Benchmarking_structure-based_three-dimensional_molecular_generative_models_using_GenBench3D_/26152900

## References

[1] Walters WP, Barzilay R. Applications of Deep Learning in Molecule Generation and Molecular Property Prediction. Acc Chem Res 2020. https://doi.org/10.1021/acs.accounts.0c00699.

[2] Bilodeau C, Jin W, Jaakkola T, Barzilay R, Jensen KF. Generative models for molecular discovery: Recent advances and challenges. WIREs Comput Mol Sci 2022;12:e1608. https://doi.org/10.1002/wcms.1608.

[3] Blaschke T, Arús-Pous J, Chen H, Margreitter C, Tyrchan C, Engkvist O, et al. REINVENT 2.0: An AI Tool for De Novo Drug Design. J Chem Inf Model 2020. https://doi.org/10.1021/acs.jcim.0c00915.

[4] Lavecchia A, Giovanni CD. Virtual Screening Strategies in Drug Discovery: A Critical Review. Curr Med Chem 2013;20:2839–60.

[5] Kitchen DB, Decornez H, Furr JR, Bajorath J. Docking and scoring in virtual screening for drug discovery: methods and applications. Nat Rev Drug Discov 2004;3:935–49. https://doi.org/10.1038/nrd1549.

[6] Hawkins PCD, Skillman AG, Nicholls A. Comparison of shape-matching and docking as virtual screening tools. J Med Chem 2007;50:74–82. https://doi.org/10.1021/jm0603365.

[7] Jin W, Barzilay R, Jaakkola T. Hierarchical Generation of Molecular Graphs using Structural Motifs 2020:arXiv:2002.03230.

[8] Liu X, Ye K, van Vlijmen HWT, Emmerich MTM, IJzerman AP, van Westen GJP. DrugEx v2: de novo design of drug molecules by Pareto-based multi-objective reinforcement learning in polypharmacology. J Cheminformatics 2021;13:85. https://doi.org/10.1186/s13321-021-00561-9.

[9] Mercado R, Rastemo T, Lindelöf E, Klambauer G, Engkvist O, Chen H, et al. Graph networks for molecular design. Mach Learn Sci Technol 2021;2:025023. https://doi.org/10.1088/2632-2153/abcf91.

[10] Thomas M, Smith RT, O'Boyle NM, De Graaf C, Bender A. Comparison of structure- and ligand-based scoring functions for deep generative models: a GPCR case study. J Cheminformatics 2021;13. https://doi.org/10.1186/s13321-021-00516-0.

[11] Baillif B, Cole J, McCabe P, Bender A. Deep generative models for 3D molecular structure. Curr Opin Struct Biol 2023;80:102566. https://doi.org/10.1016/j.sbi.2023.102566.

[12] Xie W, Wang F, Li Y, Lai L, Pei J. Advances and Challenges in De Novo Drug Design Using Three-Dimensional Deep Generative Models. J Chem Inf Model 2022;62:2269–79. https://doi.org/10.1021/acs.jcim.2c00042.

[13] Ragoza M, Masuda T, Ryan Koes D. Generating 3D molecules conditional on receptor binding sites with deep generative models. Chem Sci 2022;13:2701–13. https://doi.org/10.1039/D1SC05976A.

[14] O'Boyle NM, Banck M, James CA, Morley C, Vandermeersch T, Hutchison GR. Open Babel: An open chemical toolbox. J Cheminformatics 2011;3:33. https://doi.org/10.1186/1758-2946-3-33.

[15] Luo S, Guan J, Ma J, Peng J. A 3D Molecule Generative Model for Structure-Based Drug Design 2022. https://doi.org/10.48550/arXiv.2203.10446.

[16] Peng X, Luo S, Guan J, Xie Q, Peng J, Ma J. Pocket2Mol: Efficient Molecular Sampling Based on 3D Protein Pockets 2022. https://doi.org/10.48550/arXiv.2205.07249.

[17] Guan J, Qian WW, Peng X, Su Y, Peng J, Ma J. 3D Equivariant Diffusion for Target-Aware Molecule Generation and Affinity Prediction 2023. https://doi.org/10.48550/arXiv.2303.03543.

[18] Schneuing A, Du Y, Harris C, Jamasb A, Igashov I, Du W, et al. Structure-based Drug Design with Equivariant Diffusion Models 2023. https://doi.org/10.48550/arXiv.2210.13695.

[19] Zhang O, Zhang J, Jin J, Zhang X, Hu R, Shen C, et al. ResGen is a pocket-aware 3D molecular generation model based on parallel multiscale modelling. Nat Mach Intell 2023;5:1020–30. https://doi.org/10.1038/s42256-023-00712-7.

[20] Polykovskiy D, Zhebrak A, Sanchez-Lengeling B, Golovanov S, Tatanov O, Belyaev S, et al. Molecular Sets (MOSES): A Benchmarking Platform for Molecular Generation Models. Front Pharmacol 2020;11.

[21] Brown N, Fiscato M, Segler MHS, Vaucher AC. GuacaMol: Benchmarking Models for de Novo Molecular Design. J Chem Inf Model 2019;59:1096–108. https://doi.org/10.1021/acs.jcim.8b00839.

[22] rdkit/rdkit: 2023_09_2 (Q3 2023) Release n.d. https://doi.org/10.5281/zenodo.10099869.

[23] Ertl P, Schuffenhauer A. Estimation of synthetic accessibility score of drug-like molecules based on molecular complexity and fragment contributions. J Cheminformatics 2009;1:8. https://doi.org/10.1186/1758-2946-1-8.

[24] Bickerton GR, Paolini GV, Besnard J, Muresan S, Hopkins AL. Quantifying the chemical beauty of drugs. Nat Chem 2012;4:90–8. https://doi.org/10.1038/nchem.1243.

[25] Handa K, Thomas MC, Kageyama M, Iijima T, Bender A. On the difficulty of validating molecular generative models realistically: a case study on public and proprietary data. J Cheminformatics 2023;15:112. https://doi.org/10.1186/s13321-023-00781-1.

[26] Langevin M, Grebner C, Güssregen S, Sauer S, Li Y, Matter H, et al. Impact of Applicability Domains to Generative Artificial Intelligence. ACS Omega 2023;8:23148–67. https://doi.org/10.1021/acsomega.3c00883.

[27] Buttenschoen M, M. Morris G, M. Deane C. PoseBusters: AI-based docking methods fail to generate physically valid poses or generalise to novel sequences. Chem Sci 2024;15:3130–9. https://doi.org/10.1039/D3SC04185A.

[28] Harris C, Didi K, Jamasb AR, Joshi CK, Mathis SV, Lio P, et al. Benchmarking Generated Poses: How Rational is Structure-based Drug Design with Generative Models? 2023. https://doi.org/10.48550/arXiv.2308.07413.

[29] Groom CR, Bruno IJ, Lightfoot MP, Ward SC. The Cambridge Structural Database. Acta Crystallogr Sect B Struct Sci Cryst Eng Mater 2016;72:171–9. https://doi.org/10.1107/S2052520616003954.

[30] Cottrell SJ, Olsson TSG, Taylor R, Cole JC, Liebeschuetz JW. Validating and Understanding Ring Conformations Using Small Molecule Crystallographic Data. J Chem Inf Model 2012;52:956–62. https://doi.org/10.1021/ci200439d.

[31] McCabe P, Cole J. Assessing conformations of small molecules with crystallographic databases. J Appl Crystallogr 2023;56:420–31. https://doi.org/10.1107/S1600576723000948.

[32] Tong J, Zhao S. Large-Scale Analysis of Bioactive Ligand Conformational Strain Energy by Ab Initio Calculation. J Chem Inf Model 2021;61:1180–92. https://doi.org/10.1021/acs.jcim.0c01197.

[33] RDKit Cookbook — The RDKit 2022.09.1 documentation 2023. https://www.rdkit.org/docs/Cookbook.html (accessed October 24, 2022).

[34] Landrum G, Tosco P, Kelley B, sriniker, gedeck, Ric, et al. rdkit/rdkit: 2020_09_1 (Q3 2020) Release 2020. https://doi.org/10.5281/zenodo.4107869.

[35] Bryant MJ, Black SN, Blade H, Docherty R, Maloney AGP, Taylor SC. The CSD Drug Subset: The Changing Chemistry and Crystallography of Small Molecule Pharmaceuticals. J Pharm Sci 2019;108:1655–62. https://doi.org/10.1016/j.xphs.2018.12.011.

[36] Wishart DS, Feunang YD, Guo AC, Lo EJ, Marcu A, Grant JR, et al. DrugBank 5.0: a major update to the DrugBank database for 2018. Nucleic Acids Res 2018;46:D1074–82. https://doi.org/10.1093/nar/gkx1037.

[37] McCabe P, Korb O, Cole J. Kernel Density Estimation Applied to Bond Length, Bond Angle, and Torsion Angle Distributions. J Chem Inf Model 2014;54:1284–8. https://doi.org/10.1021/ci500156d.

[38] Pedregosa F, Varoquaux G, Gramfort A, Michel V, Thirion B, Grisel O, et al. Scikit-learn: Machine Learning in Python. J Mach Learn Res 2011;12:2825–30.

[39] Harris CR, Millman KJ, van der Walt SJ, Gommers R, Virtanen P, Cournapeau D, et al. Array programming with NumPy. Nature 2020;585:357–62. https://doi.org/10.1038/s41586-020-2649-2.

[40] Schulz-Gasch T, Schärfer C, Guba W, Rarey M. TFD: Torsion Fingerprints As a New Measure To Compare Small Molecule Conformations. J Chem Inf Model 2012;52:1499–512. https://doi.org/10.1021/ci2002318.

[41] Cole JC, Korb O, Mccabe P, Read MG, Taylor R. Knowledge-Based Conformer Generation Using the Cambridge Structural Database. J Chem Inf Model 2018;58:615–29. https://doi.org/10.1021/acs.jcim.7b00697.

[42] Halgren TA. MMFF VI. MMFF94s option for energy minimization studies. J Comput Chem 1999;20:720–9. https://doi.org/10.1002/(SICI)1096-987X(199905)20:7<720::AID-JCC7>3.0.CO;2-X.

[43] Francoeur PG, Masuda T, Sunseri J, Jia A, Iovanisci RB, Snyder I, et al. Three-Dimensional Convolutional Neural Networks and a Cross-Docked Data Set for Structure-Based Drug Design. J Chem Inf Model 2020;60:4200–15. https://doi.org/10.1021/acs.jcim.0c00411.

[44] Gowers RJ, Linke M, Barnoud J, Reddy TJE, Melo MN, Seyler SL, et al. MDAnalysis: A Python Package for the Rapid Analysis of Molecular Dynamics Simulations. Proc 15th Python Sci Conf 2016:98–105. https://doi.org/10.25080/Majora-629e541a-00e.

[45] Virtanen P, Gommers R, Oliphant TE, Haberland M, Reddy T, Cournapeau D, et al. SciPy 1.0: fundamental algorithms for scientific computing in Python. Nat Methods 2020;17:261–72. https://doi.org/10.1038/s41592-019-0686-2.

[46] Eberhardt J, Santos-Martins D, Tillack AF, Forli S. AutoDock Vina 1.2.0: New Docking Methods, Expanded Force Field, and Python Bindings. J Chem Inf Model 2021;61:3891–8. https://doi.org/10.1021/acs.jcim.1c00203.

[47] Ravindranath PA, Forli S, Goodsell DS, Olson AJ, Sanner MF. AutoDockFR: Advances in Protein-Ligand Docking with Explicitly Specified Binding Site Flexibility. PLOS Comput Biol 2015;11:e1004586. https://doi.org/10.1371/journal.pcbi.1004586.

[48] Meeko: preparation of small molecules for AutoDock 2023.

[49] Trott O, Olson AJ. AutoDock Vina: Improving the speed and accuracy of docking with a new scoring function, efficient optimization, and multithreading. J Comput Chem 2009:455–61. https://doi.org/10.1002/jcc.21334.

[50] Friesner RA, Banks JL, Murphy RB, Halgren TA, Klicic JJ, Mainz DT, et al. Glide: A New Approach for Rapid, Accurate Docking and Scoring. 1. Method and Assessment of Docking Accuracy. J Med Chem 2004;47:1739–49. https://doi.org/10.1021/jm0306430.

[51] Schrödinger Release 2023-4: Maestro, Schrödinger, LLC, New York, NY, 2023. n.d.

[52] Verdonk ML, Cole JC, Hartshorn MJ, Murray CW, Taylor RD. Improved protein-ligand docking using GOLD. Proteins Struct Funct Bioinforma 2003;52:609–23. https://doi.org/10.1002/prot.10465.

[53] Verkhivker GM, Bouzida D, Gehlhaar DK, Rejto PA, Arthurs S, Colson AB, et al. Deciphering common failures in molecular docking of ligand-protein complexes. J Comput Aided Mol Des 2000;14:731–51. https://doi.org/10.1023/A:1008158231558.

[54] Benjamini Y, Hochberg Y. Controlling the False Discovery Rate: A Practical and Powerful Approach to Multiple Testing. J R Stat Soc Ser B Methodol 1995;57:289–300.

[55] Liebeschuetz J, Hennemann J, Olsson T, Groom CR. The good, the bad and the twisted: a survey of ligand geometry in protein crystal structures. J Comput Aided Mol Des 2012;26:169–83. https://doi.org/10.1007/s10822-011-9538-6.

[56] Liebeschuetz JW. The Good, the Bad, and the Twisted Revisited: An Analysis of Ligand Geometry in Highly Resolved Protein–Ligand X-ray Structures. J Med Chem 2021;64:7533–43. https://doi.org/10.1021/acs.jmedchem.1c00228.

[57] Sitzmann M, Weidlich IE, Filippov IV, Liao C, Peach ML, Ihlenfeldt W-D, et al. PDB Ligand Conformational Energies Calculated Quantum-Mechanically. J Chem Inf Model 2012;52:739–56. https://doi.org/10.1021/ci200595n.

[58] Koes DR, Baumgartner MP, Camacho CJ. Lessons Learned in Empirical Scoring with smina from the CSAR 2011 Benchmarking Exercise. J Chem Inf Model 2013;53:1893–904. https://doi.org/10.1021/ci300604z.

[59] King-Smith E. Transfer learning for a foundational chemistry model. Chem Sci 2024;15:5143–51. https://doi.org/10.1039/D3SC04928K.

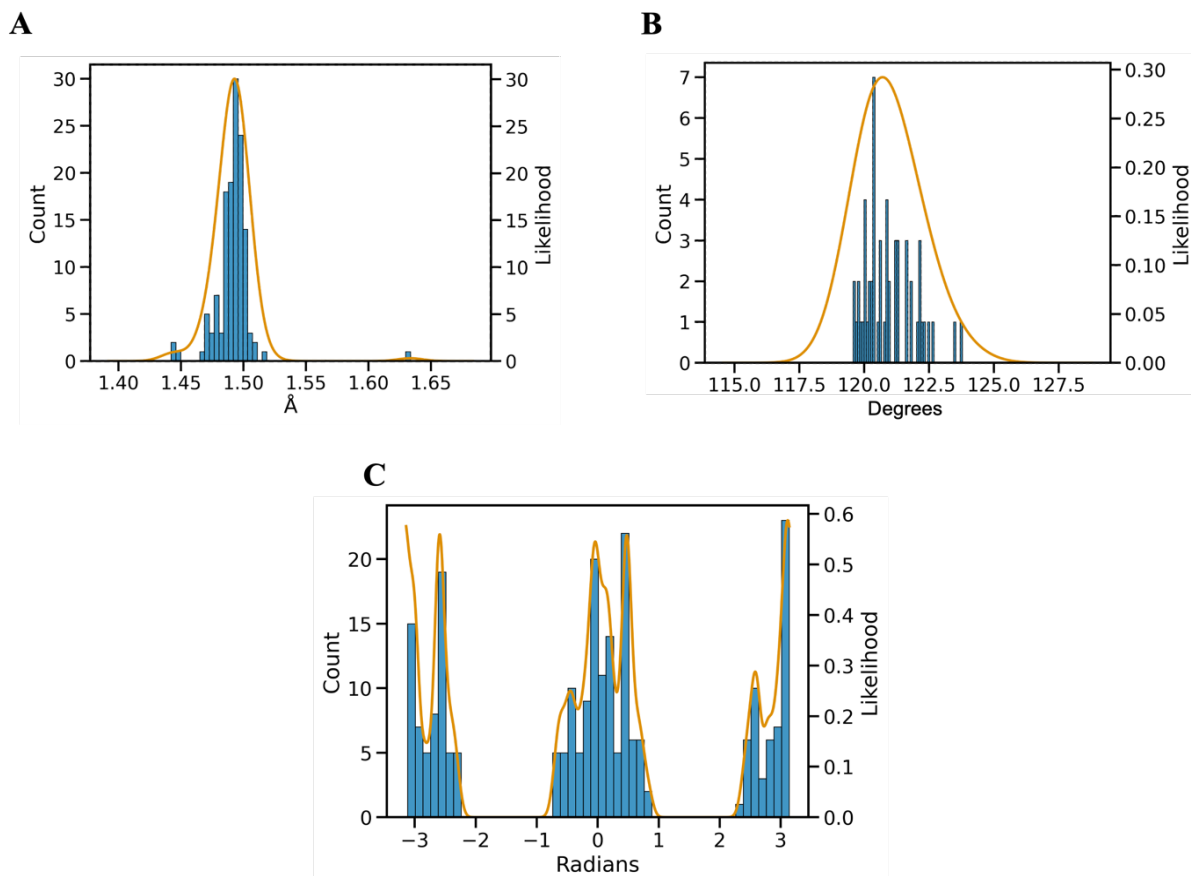

*Figure S 1: Distributions of bond lengths (A), valence angles (B) and torsion angles (C) for the patterns displayed in Figure 1. The histograms of values (counts) are shown with blue bars, while the likelihoods evaluated by kernel density estimation are shown in yellow lines.*

## Evaluation of molecular graph metrics

Generated molecules have validity above 0.70, uniqueness above 0.89, novelty above 0.98 and diversity above 0.91, which are suitable values for the common objective of models to generate new, valid and diverse molecules. The median maximum training similarity range from 0.17 for DiffSBDD to 0.27 for ResGen and the median molecular weight is smaller than the CrossDocked ligands (353 g/mol) for LiGAN (224 g/mol), 3D-SBDD (243 g/mol), Pocket2Mol (238 g/mol) and ResGen (258 g/mol), and similar for TargetDiff (349 g/mol) and DiffSBDD (331 g/mol). Generated molecules have similar median QED close to 0.5 except for Pocket2Mol and ResGen with a higher median QED around 0.6. The SAScore, estimating synthetic accessibility (ranging between 0 and 10, lower is better), was around 4.9 for molecules generated by LiGAN, 3D-SBDD, TargetDiff and DiffSBDD, while it was close to the CrossDocked ligand (3.1) for Pocket2Mol and ResGen with 2.9 and 2.8 respectively. Based on these numbers, the evaluated models seem to show performances on a par with those from SMILES or graph generators evaluated in GuacaMol and MOSES. The molecules with best estimated synthetic accessibility, drug-likeness, and with ring proportion close to training ligands are Pocket2Mol and ResGen (quantify), that are both based on the same auto-regressive architecture, but they generate molecules in general smaller to training ligands. We also analysed the ring size distribution, reported in Table S1. We found that there is a high fraction of 3-membered rings for LiGAN (0.60 ), 3D-SBDD (0.47) and DiffSBDD (0.31) compared to CrossDocked ligands (0.01), a high fraction of 4-membered rings for LiGAN (0.15) and a high fraction of 7-membered rings for TargetDiff (0.11) compared to CrossDocked ligands (0.01). This indicates that the generated chemical space is quite different, as pointed earlier by the low maximum similarity to CrossDocked training set.

*Table S 1: Molecular graph based metrics*

| Molecule set | Validity$_G$ | Uniqueness$_G$ | Diversity$_G$ | Novelty$_G$ | Median Maximum training similarity | Median MW | Median logP | Median SAScore | Median QED |
|---|---|---|---|---|---|---|---|---|---|
| LiGAN | 0.753 | 0.993 | 0.923 | 0.998 | 0.182 | 224.344 | 0.456 | 4.920 | 0.461 |
| 3D-SBDD | 0.815 | 0.938 | 0.936 | 0.993 | 0.183 | 243.215 | -0.091 | 4.861 | 0.460 |
| Pocket2Mol | 0.982 | 0.901 | 0.921 | 0.985 | 0.257 | 239.227 | 1.252 | 2.910 | 0.591 |
| TargetDiff | 1.000 | 0.987 | 0.925 | 0.996 | 0.178 | 350.376 | 1.386 | 4.821 | 0.483 |
| DiffSBDD | 1.000 | 0.998 | 0.922 | 0.999 | 0.168 | 331.500 | 2.392 | 4.993 | 0.470 |
| ResGen | 1.000 | 0.893 | 0.917 | 0.979 | 0.269 | 262.329 | 1.443 | 2.816 | 0.607 |
| CrossDocked training set | 1.000 | 0.088 | 0.925 | NA | NA | 353.413 | 2.392 | 3.184 | 0.532 |
| CSD Drug | 0.979 | 0.315 | 0.941 | NA | NA | 266.213 | 1.077 | 3.657 | 0.555 |

*Table S 2: Ring proportions in generated molecules*

| Model | Ring proportion | | | | | |
|---|---|---|---|---|---|---|
| | 3 | 4 | 5 | 6 | 7 | > 7 |
| LiGAN | 0.603 | 0.150 | 0.134 | 0.080 | 0.015 | 0.018 |
| 3D-SBDD | 0.468 | 0.002 | 0.139 | 0.338 | 0.022 | 0.032 |
| Pocket2Mol | 0.001 | 0.000 | 0.203 | 0.762 | 0.024 | 0.010 |
| TargetDiff | 0.000 | 0.027 | 0.297 | 0.490 | 0.117 | 0.069 |
| DiffSBDD | 0.314 | 0.048 | 0.192 | 0.324 | 0.085 | 0.038 |
| ResGen | 0.000 | 0.000 | 0.196 | 0.781 | 0.013 | 0.009 |
| CrossDocked training set | 0.015 | 0.003 | 0.272 | 0.696 | 0.008 | 0.007 |
| CSD subset | 0.008 | 0.011 | 0.184 | 0.713 | 0.016 | 0.068 |

*Table S 3: Median conformation quality metrics on the set of relaxed molecules*

| Model | Validity3D | Minimum q-value | | | Geometric mean of q-values | | | Strain energy (kcal/mol) |
|---|---|---|---|---|---|---|---|---|
| | | Bond | Angle | Bond and angle | Bond | Angle | Bond and angle | |
| LiGAN | 0.551 | 0.039 | 0.005 | 0.003 | 0.369 | 0.165 | 0.249 | 30.0 |
| 3D_SBDD | 0.703 | 0.080 | 0.018 | 0.015 | 0.520 | 0.312 | 0.399 | 15.7 |
| Pocket2Mol | 0.783 | 0.079 | 0.024 | 0.016 | 0.569 | 0.439 | 0.481 | 10.1 |
| TargetDiff | 0.544 | 0.029 | 0.003 | 0.002 | 0.457 | 0.257 | 0.331 | 28.7 |
| DiffSBDD | 0.532 | 0.046 | 0.002 | 0.002 | 0.483 | 0.215 | 0.318 | 36.2 |
| ResGen | 0.687 | 0.055 | 0.012 | 0.008 | 0.528 | 0.362 | 0.420 | 14.3 |

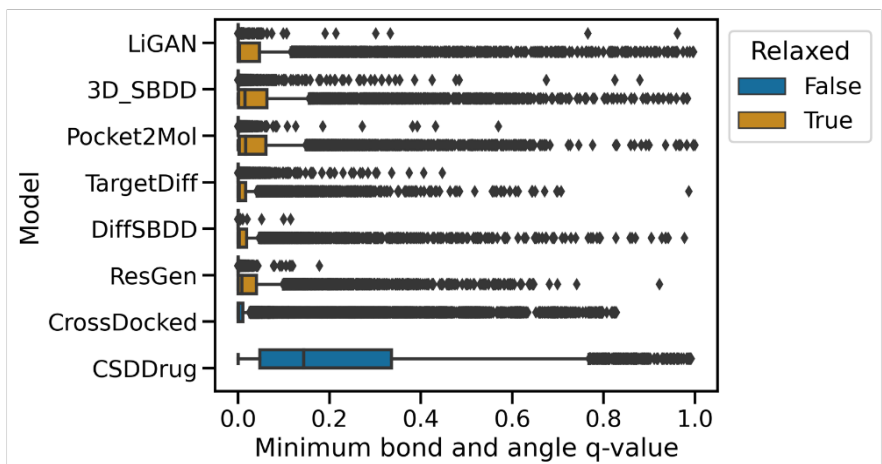

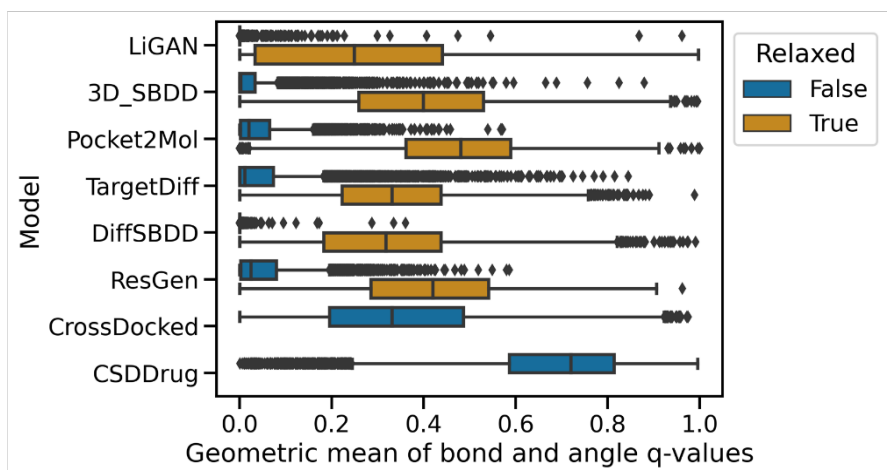

*Figure S 2: Distribution of the geometric means (A) and minima (B) q-values for bond and valence angle patterns in molecules generated by various models*

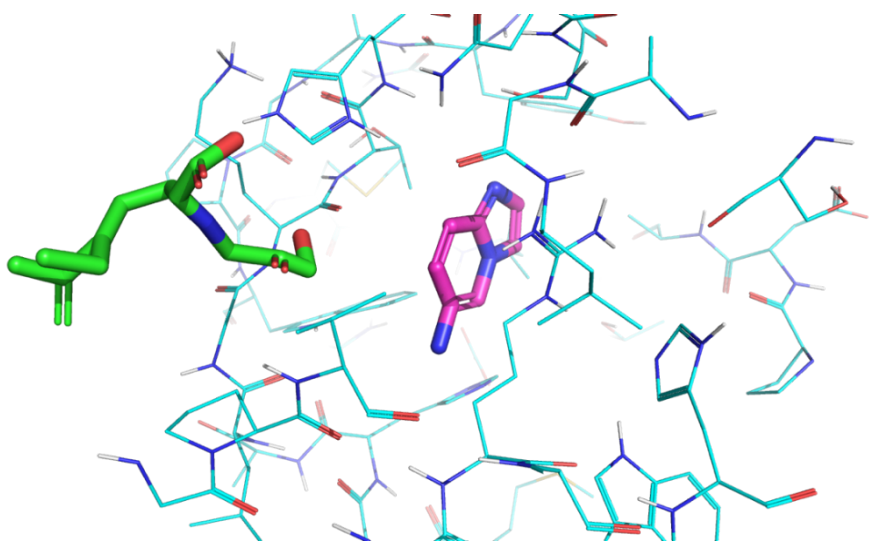

*Figure S 3: Molecule generated by DiffSBDD. The generated molecule is shown in liquorice representation with green carbon atoms, the native ligand is shown in liquorice representation with purple carbon atoms, while the protein pocket is shown in wire representation in cyan carbon atoms.*

*Table S 4: Conformation difference metrics on the set of valid relaxed molecules. \*: for CrossDocked and CSD Drug, no relaxation was performed, and the Novelty3D was not evaluated as they are reference sets.*

| Molecule set | Uniqueness$_{3D}$ | Diversity$_{3D}$ | Novelty$_{3D}$ |
|---|---|---|---|
| LiGAN | 0.800 | 0.488 | 0.333 |
| 3D_SBDD | 0.429 | 0.250 | 0.402 |
| Pocket2Mol | 0.382 | 0.199 | 0.121 |
| TargetDiff | 0.683 | 0.376 | 0.272 |
| DiffSBDD | 0.625 | 0.274 | 0.500 |
| ResGen | 0.419 | 0.264 | 0.170 |
| CrossDocked (training)* | 0.096 | 0.086 | NA |
| CSD Drug* | 0.193 | 0.104 | NA |

*Table S 5: Pocket-based metrics: fractions of generated molecules out of the pocket*

| Model | Distance to native centroid > 10 Å |
|---|---|
| LiGAN | 0 |
| 3D-SBDD | 0.001 |
| Pocket2Mol | 0 |
| TargetDiff | 0.001 |
| DiffSBDD | 0.057 |
| ResGen | 0 |

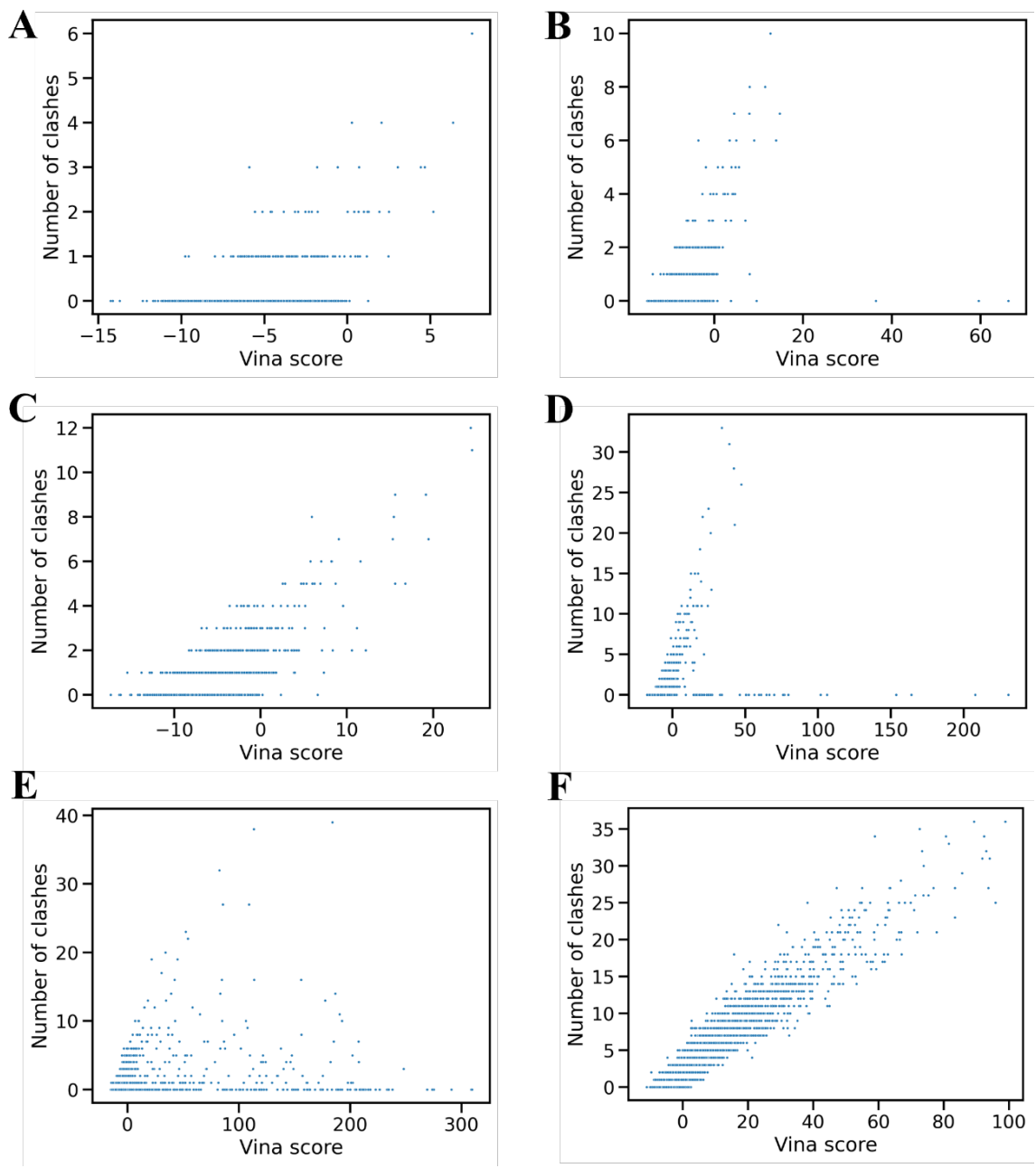

*Figure S 4: Relationship between the Vina score and the number of clashes for the set of valid relaxed molecules generated by LiGAN (A), 3D-SBDD (B), Pocket2Mol (C), TargetDiff (D), DiffSBDD (E) and ResGen (F)*

# Relationship between equilibrium constant and estimated binding free energy

The Vina scoring function estimates the binding affinity by its free energy ΔG in kcal·mol$^{-1}$. The equilibrium constant $K$ (that can be measured in binding assay in the form of a dissociation constant Kd or inhibitory constant Ki, in mol·L$^{-1}$) is directly correlated to ΔG, usually computed in J·mol$^{-1}$ (1 kcal = 4184 J), via the following equation:

$$\Delta G = -R \cdot T \cdot \ln(K)$$

where R is the molar gas constant (8.314 J·Kelvin$^{-1}$·mol$^{-1}$), and T the temperature in Kelvin. At room temperature T = 295 Kelvin, a ten-fold difference between two values of $K$ leads to a difference in ΔG in kcal/mol equal to 1.3; a two-fold difference in $K$ is a ΔG difference of 0.4.

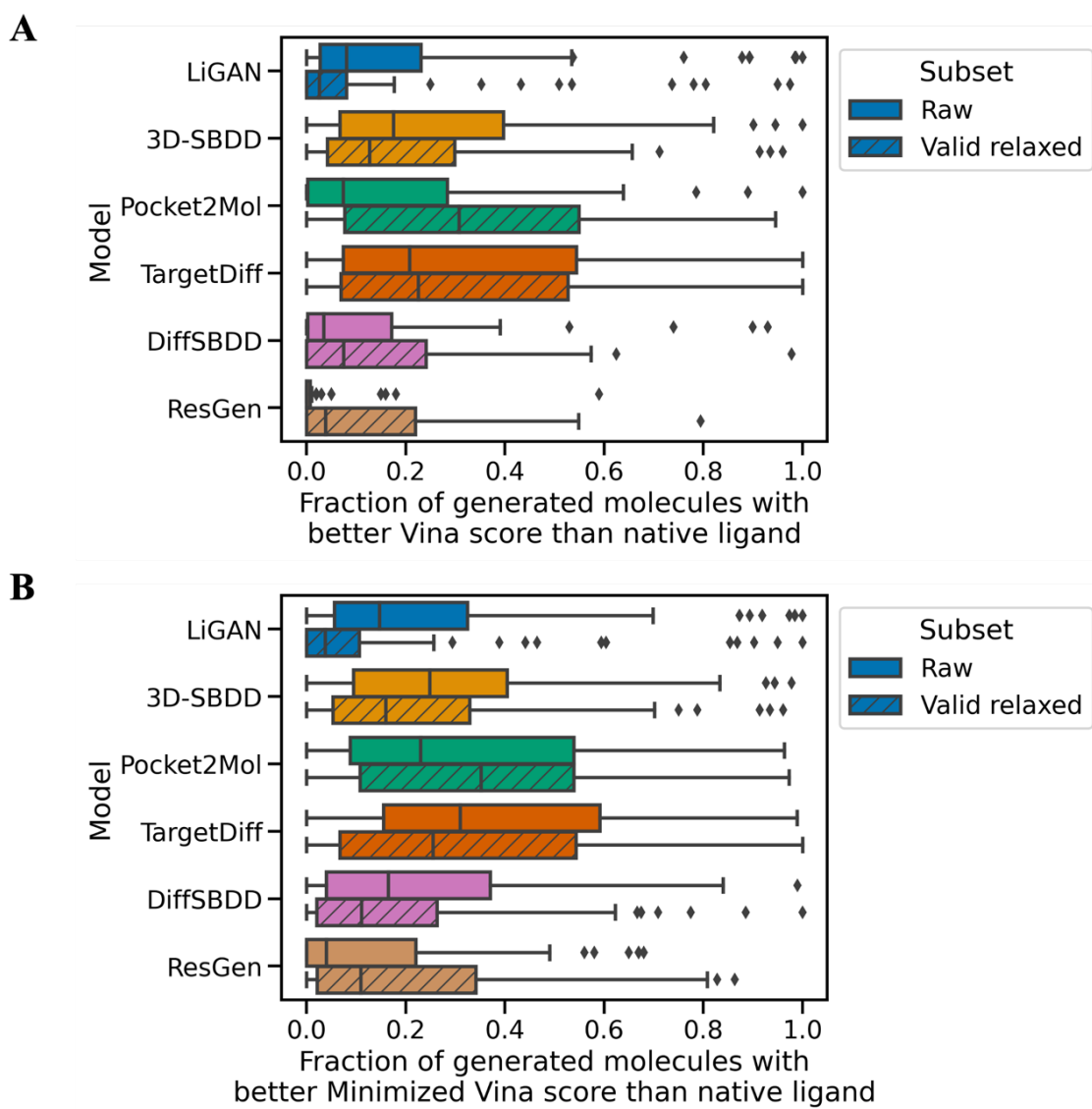

*Figure S 5: Distributions of fractions of generated molecules with a better Vina (A) or minimized Vina (B) score than the native ligand, for the sets of raw (open bars) and valid relaxed (hatched bars) molecules.*

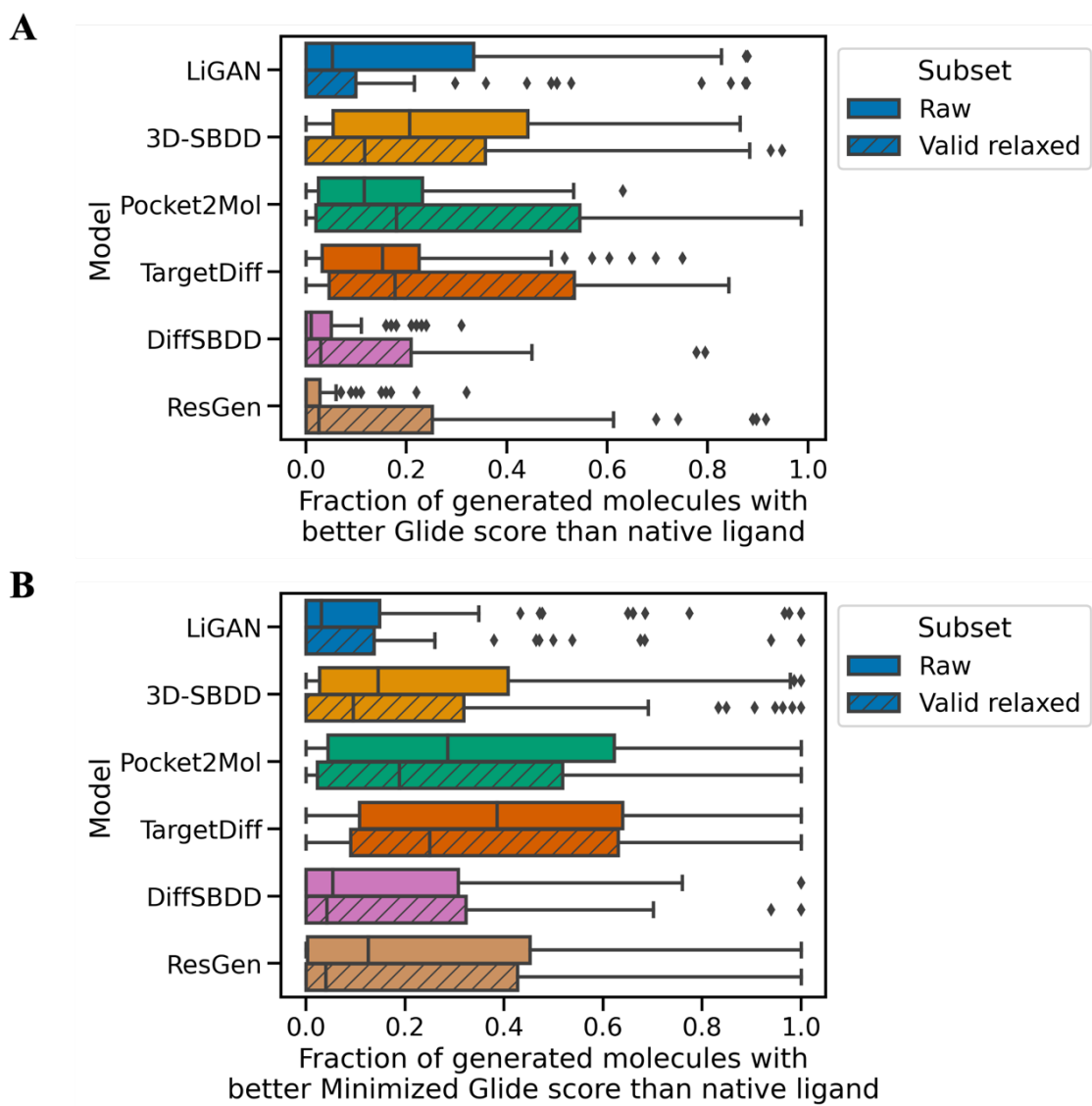

*Figure S 6: Distributions of fractions of generated molecules with a better Glide (A) or minimized Glide (B) score than the native ligand, for the sets of raw (open bars) and valid relaxed (hatched bars) molecules.*

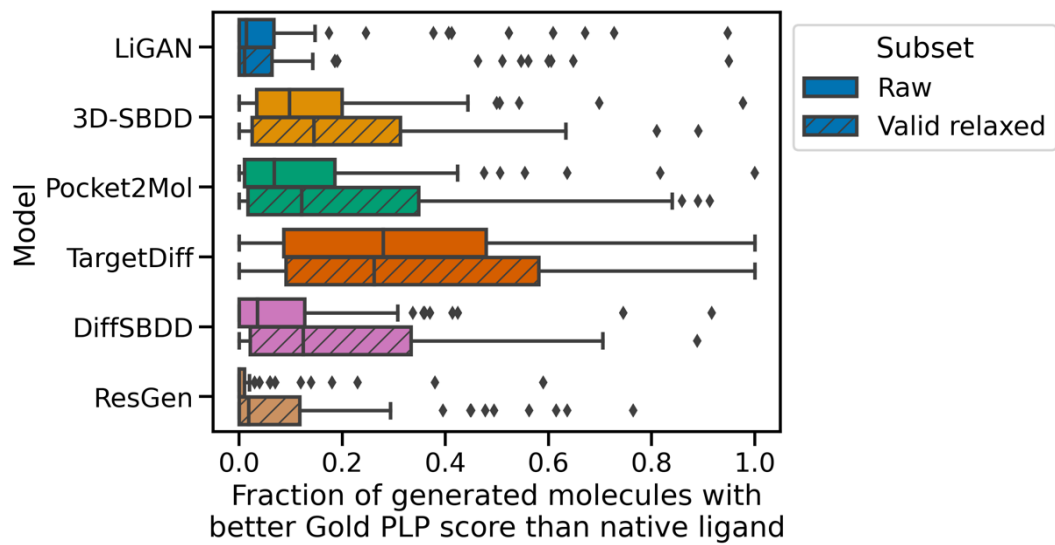

*Figure S 7: Distributions of fractions of generated molecules with a better Gold PLP score than the native ligand, for the sets of raw (open bars) and valid relaxed (hatched bars) molecules.*